\documentclass[journal, final, onecolumn,12pt]{IEEEtran}  
\usepackage{cite}
\usepackage{graphicx}
\usepackage{subfigure}
\usepackage{mathrsfs}
\usepackage{amsmath}
\usepackage{amsfonts}
\usepackage{amssymb}

\newtheorem{theorem}{{Theorem}}
\newtheorem{lemma}[theorem]{{Lemma}}
\newtheorem{corollary}[theorem]{{Corollary}}
\newtheorem{definition}{{Definition}}
\newtheorem{proposition}[theorem]{{Proposition}}

\DeclareMathOperator*{\argmin}{argmin}

\newcommand{\mb}{\mathbf}
\newcommand{\qed}{\hspace*{\fill} $\Box$ \\}

\begin{document}

\title{Connectivity, Percolation, and Information Dissemination in Large-Scale Wireless Networks with Dynamic Links}

\author{Zhenning~Kong,~\IEEEmembership{Student Member,~IEEE,}
        Edmund~M. Yeh,~\IEEEmembership{Member,~IEEE,}\\
\thanks{This research is supported in part by National
Science Foundation (NSF) Cyber Trust grant CNS-0716335, and by Army Research Office (ARO)
grant W911NF-07-1-0524.}
\thanks{The material in this paper was presented in part at the 27th IEEE Conference on Computer
Communications (INFOCOM), Phoenix, AZ, April 2008.}
\thanks{Z. Kong and Edmund~M. Yeh are with the Department of Electrical Engineering, Yale University
(email: zhenning.kong@yale.edu, edmund.yeh@yale.edu)}}

\markboth{Submitted to \emph{IEEE Transactions on Information Theory}}{Submitted to
\emph{IEEE Transactions on Information Theory}}

\maketitle

\begin{abstract}
We investigate the problem of disseminating broadcast messages in wireless networks with
time-varying links from a percolation-based perspective.  Using a model of wireless
networks based on random geometric graphs with dynamic on-off links, we show that the
delay for disseminating broadcast information exhibits two behavioral regimes,
corresponding to the phase transition of the underlying network connectivity. When the
dynamic network is in the subcritical phase, ignoring propagation delays, the delay
scales linearly with the Euclidean distance between the sender and the receiver. When the
dynamic network is in the supercritical phase, the delay scales sub-linearly with the
distance. Finally, we show that in the presence of a non-negligible propagation delay,
the delay for information dissemination scales linearly with the Euclidean distance in
both the subcritical and supercritical regimes, with the rates for the linear scaling
being different in the two regimes.
\end{abstract}

\baselineskip 22 pt

\section{Introduction}

Large-scale wireless networks for the gathering, processing, and dissemination of
information have become an important part of modern life.  To ensure that important
broadcast messages can be received by each node in a wireless network, the network needs
to maintain full connectivity~\cite{GuKu98}.  Here, the system ensures that each pair of
network nodes are connected by a path of consecutive links. In large-scale wireless
networks exposed to severe natural hazards, enemy attacks, and resource depletion,
however, the full connectivity criterion may be overly restrictive or impossible to
achieve.  In these challenging environments, the system designer may reasonably aim for a
slightly weaker notion of connectivity, one which ensures that a high fraction of the
network nodes can successfully receive broadcast messages.  This latter viewpoint can be
explored using the mathematical theory of percolation~\cite{Gi61, Gr99, MeRo96, Pe03}.

In this paper, we investigate the problem of information dissemination in wireless
networks from a percolation-based perspective.  Using a model of wireless networks based
on random geometric graphs with dynamic on-off links, we show that the delay for
disseminating broadcast information exhibits a phase transition as a function of the
underlying node density.  Assuming zero propagation delay, we show that in the
subcritical regime, the delay scales linearly with the distance between the sender and
receiver. In the supercritical regime, the delay scales sub-linearly with the distance.

In recent years, percolation theory, especially continuum percolation theory~\cite{
MeRo96, Pe03}, has become a useful tool for the analysis of large-scale wireless
networks~\cite{BoNrFrMe03, FrBoCoBrMe05, DoMaTh04, DoFrTh05, DoBaTh05, DoFrMaMeTh06,
FrDoTsTh07, KoYe07-4, KoYe08-1, KoYe08-2}.  A major focus of continuum percolation theory
is the random geometric graph in which nodes are distributed according to a Poisson point
process with constant density $\lambda$, and two nodes share a link if they are within
distance 1 of each other. A fundamental result of continuum percolation concerns a phase
transition effect whereby the macroscopic behavior of the random geometric graph is very
different for densities below and above the critical density $\lambda_c$. For $\lambda <
\lambda_c$ (subcritical), the connected component containing the origin contains a finite
number of points almost surely. For $\lambda> \lambda_c$ (supercritical), the connected
component containing the origin contains an infinite number of points with a positive
probability~\cite{Gr99, MeRo96, Pe03}.

Wireless networks are subject to multi-user interference, fading, and noise.  Thus, even
when two nodes are within each other's transmission range, a viable communication link
may not exist~\cite{FrBoCoBrMe05}.   Furthermore, due to fading, the link quality can
vary dynamically in time, inducing a frequently changing network topology.  To capture
these effects, we model a wireless network by a random geometric graph in which each
link's functionality (activity) varies dynamically in time according to a Markov on-off
process.  Using this model, we investigate the problem of disseminating broadcast
messages in wireless networks.  Due to the dynamic on-off behavior of links, a delay is
incurred in transmitting a broadcast message from the sender to the receiver even when
propagation delay is ignored.   The main question we address is how this delay scales
with the distance between the sender and the receiver.

As a first step, we show that the connectivity of the network with dynamic links exhibits
a phase transition as a function of the underlying node density.  We characterize the
critical density for this phase transition in terms of the link state process.  Next, we
show that the delay for disseminating broadcast information exhibits two behavioral
regimes, corresponding to the phase transition of the underlying network connectivity.
When the dynamic network is in the subcritical phase, ignoring propagation delays, the
delay scales linearly with the Euclidean distance between the sender and the receiver.
This follows from the fact that in this regime, connectivity decays exponentially with
distance, and on average, any information dissemination process is blocked by inactive
links after the message travels a finite distance (and is resumed after the next link
turns back on).  When the dynamic network is in the supercritical phase, the delay scales
sub-linearly with the distance between the sender and the receiver.  In this case, the
delay is determined largely by the amount of time it takes for the message to reach the
infinite connected component of the dynamic network.  Finally, we characterize the delay
for information dissemination when propagation delays are taken into account.  Here, the
problem becomes more subtle. We show that, with the presence of a non-negligible
propagation delay, the delay for information dissemination scales linearly with the
Euclidean distance between the sender and the receiver in both the subcritical and
supercritical regimes, with the rates for the linear scaling being different in the two
regimes.

In order to study the behavior of information dissemination delay in wireless networks
with dynamic links, we model the problem as a first passage percolation
process~\cite{Ke87, De03}. Similar first passage percolation problems have been studied
within the context of lattices~\cite{Gr99, Ke87}. Related continuum models are considered
in~\cite{De03, DoMaTh04, KoYe07-4}. In~\cite{De03}, Deijfen studies a continuum growth
model for a spreading infection with Poisson point processes, and shows that the shape of
the infected cluster scales linearly with time in all directions. In~\cite{DoMaTh04},
Dousse \emph{et al.} study how the latency of information dissemination scales within an
independent site percolation model in wireless sensor networks. There, each sensor
independently switches between the on and off states at random from time to time. The
authors show that the latency scales linearly with the distance between the sender and
the receiver when the dynamic sensor network is in the subcritical phase.
In~\cite{KoYe07-4}, the authors obtain similar results for degree-dependent site
percolation model in wireless sensor networks. Unlike the problems studied
in~\cite{DoMaTh04, KoYe07-4}, however, the problem addressed in this paper requires a
bond percolation model, which demands different modelling and analysis techniques.
Furthermore, in contrast to~\cite{De03, DoMaTh04}, we also study the delay scaling for
networks in the supercritical phase.  Finally, we present new results regarding networks
with propagation delay.

The remainder of this paper is organized as follows. In Section II, we outline some
preliminary results for random geometric graphs and continuum percolation. In Section
III, we present a simple model for wireless networks with static unreliable links.  In
Section IV, we introduce a more sophisticated model for wireless networks with dynamic
unreliable links, and present our main results regarding percolation-based connectivity
and information dissemination within this model. In Section V, we present simulation
results, and finally, in Section VI, we conclude the paper.

\section{Random Geometric Graphs and Continuum Percolation}

\subsection{Random Geometric Graphs}

We use random geometric graphs to model wireless networks. That is, we assume that the
network nodes are randomly placed over some area or volume, and a communication link
exists between two (randomly placed) nodes if the distance between them is sufficiently
small, so that the received power is large enough for successful decoding. A mathematical
model for this is as follows. Let $\|\cdot\|$ be the Euclidean norm, and $f(\cdot)$ be
some probability density function (p.d.f.) on $\mathbb{R}^d$. Let ${\mb X}_1, {\mb X}_2,
..., {\mb X}_n$ be independent and identically distributed (i.i.d.) $d$-dimensional
random variables with common density $f(\cdot)$, where ${\mb X}_i$ denotes the random
location of node $i$ in $\mathbb{R}^d$. The ensemble of graphs with undirected links
connecting all those pairs $\{{\mb x}_i, {\mb x}_j\}$ with $\|{\mb x}_i- {\mb x}_j \|\leq
r, r>0,$ is called a \emph{random geometric graph}~\cite{Pe03}, denoted by $G({\cal X}_n,
r)$. The parameter $r$ is called the characteristic radius.

In the following, we consider random geometric graphs $G({\cal X}_n, r)$ in
$\mathbb{R}^2$, with ${\mb X}_1, {\mb X}_2, ..., {\mb X}_n$ distributed i.i.d. according
to a uniform distribution in a square area ${\cal A}=[0,\sqrt{\frac{n}{\lambda}}]^2$. Let
$A=|{\cal A}|$ be the area of ${\cal A}$. There exists a link between two nodes $i$ and
$j$ if and only if $i$ lies within a circle of radius $r$ around ${\mb x}_j$. As $n$ and
$A$ both become large with the ratio $\frac{n}{A}=\lambda$ kept constant, $G({\cal X}_n,
r)$ converges in distribution to an (infinite) random geometric graph
$G(\mathcal{H}_{\lambda},r)$ induced by a homogeneous Poisson point process with density
$\lambda>0$. Due to the scaling property of random geometric graphs~\cite{MeRo96,Pe03},
we focus on $G(\mathcal{H}_{\lambda},1)$ in the following.

\subsection{Critical Density for Continuum Percolation}

To intuitively understand percolation processes in large-scale wireless networks,
consider the following example. Suppose a set of nodes are uniformly and independently
distributed at random over an area. All nodes have the same transmission radius, and two
nodes within a transmission radius of each other are assumed to communicate directly. At
first, the nodes are distributed according to a very small density. This results in
isolation and no communication among nodes. As the density increases, some clusters in
which nodes can communicate with one another directly or indirectly (via multi-hop relay)
emerge, though the sizes of these clusters are still small compared to the whole network.
As the density continues to increase, at some critical point a huge cluster containing a
large portion of the network forms. This phenomenon of a sudden and drastic change in the
global structure is called a \emph{phase transition}. The density at which phase
transition takes place is called the \emph{critical density}\cite{Gr99, MeRo96, Pe03}.

More formally, let $\mathcal{H}_{\lambda,\mathbf{0}}=\mathcal{H}_{\lambda}\cup
\{\mathbf{0}\}$, i.e., the union of the origin and the infinite homogeneous Poisson point
process with density $\lambda$. Note that in a random geometric graph induced by a
homogeneous Poisson point process, the choice of the origin can be arbitrary. We have the
following definition~\cite{MeRo96}.

\vspace{+0.1in}%
\begin{definition} For $G(\mathcal{H}_{\lambda,\mathbf{0}},1)$, let $W_{\mathbf{0}}$ be
the connected component of $G(\mathcal{H}_{\lambda,\mathbf{0}},1)$ containing
$\mathbf{0}$. Define the following critical densities:
\begin{eqnarray}
\lambda_{\#}&\triangleq&\inf \{\lambda: \Pr(|W_{\mathbf{0}}|=\infty)>0\},\\
\lambda_N&\triangleq&\inf \{\lambda: E[|W_{\mathbf{0}}|]=\infty\},\\
\lambda_c&\triangleq&\inf \{\lambda: \Pr(d(W_{\mathbf{0}})=\infty)>0\},\\
\lambda_D&\triangleq&\inf \{\lambda: E[d(W_{\mathbf{0}})]=\infty\},
\end{eqnarray}
where $|W_{\mathbf{0}}|$ is the cardinality---the number of nodes---of $W_{\mathbf{0}}$,
and $d(W_{\mathbf{0}})\triangleq\sup\{||\mathbf{x}-\mathbf{y}||: \mathbf{x},\mathbf{y}\in
W_{\mathbf{0}}\}$.
\end{definition}
\vspace{0.1in}%

As shown in Theorem 3.4 and Theorem 3.5 in~\cite{MeRo96}, these four critical densities
are identical. According to the theory of continuum percolation~\cite{MeRo96},
$0<\lambda_c<\infty$. Furthermore, when $\lambda>\lambda_c$, there exists a unique
infinite component in $G(\mathcal{H}_{\lambda,\mathbf{0}},1)$ with probability 1, and
when $\lambda<\lambda_c$, there is no infinite component in
$G(\mathcal{H}_{\lambda,\mathbf{0}},1)$ with probability 1~\cite{MeRo96}.

\section{Wireless Networks with Static Unreliable Links}

Random geometric graphs are good simplified models for wireless networks. However, due to
noise, fading,  and interference, wireless communication links between two nodes are
usually unreliable. We first use the bond percolation model on random geometric graphs to
study percolation-based connectivity of large-scale wireless networks with static
unreliable links. Given a random geometric graph $G(\mathcal{H}_{\lambda}, 1)$, let each
link of $G(\mathcal{H}_{\lambda}, 1)$ be active (independent of all other links) with
probability $p_e(d)$ which may depend on $d$, where $d = \| {\mathbf x}_i - {\mathbf
x}_j\| \leq 1$ is the length of the link $(i,j)$.  The resulting graph consisting of all
active links and their end nodes is denoted by $G(\mathcal{H}_{\lambda}, 1, p_e(\cdot))$.
This model is a specific example of the \emph{random connection model} in continuum
percolation theory~\cite{MeRo96}. In this simple model, all links in the network are
either active (on) or inactive (off) for all time. Later in this paper, we will study a
more sophisticated model where links dynamically switch between active and inactive
states from time to time.

\vspace{+0.1in}%
\begin{definition} For $G(\mathcal{H}_{\lambda,\mathbf{0}},1,p_e(\cdot))$, let $W_{\mathbf{0}}'$ be
the connected component of $G(\mathcal{H}_{\lambda,\mathbf{0}},1,p_e(\cdot))$ containing
$\mathbf{0}$. We define four critical densities:
\begin{eqnarray}
\lambda_{\#}(p_e(\cdot))&\triangleq&\inf \{\lambda: \Pr(|W_{\mathbf{0}}'|=\infty)>0\},\\
\lambda_N(p_e(\cdot))&\triangleq&\inf \{\lambda: E[|W_{\mathbf{0}}'|]=\infty\},\\
\lambda_c(p_e(\cdot))&\triangleq&\inf \{\lambda: \Pr(d(W_{\mathbf{0}}')=\infty)>0\},\\
\lambda_D(p_e(\cdot))&\triangleq&\inf \{\lambda: E[d(W_{\mathbf{0}}')]=\infty\},
\end{eqnarray}
where $|W_{\mathbf{0}}'|$ is the cardinality---the number of nodes---of
$W_{\mathbf{0}}'$, and $d(W_{\mathbf{0}}')\triangleq\sup\{||\mathbf{x}-\mathbf{y}||:
\mathbf{x},\mathbf{y}\in W_{\mathbf{0}}'\}$.
\end{definition}
\vspace{+0.1in}%

As in traditional continuum percolation, the following proposition asserts that the above
four critical densities are identical.

\vspace{0.1in}%
\begin{proposition}\label{Proposition-Densities}
For $G(\mathcal{H}_{\lambda,\mathbf{0}},1,p_e(\cdot))$, we have
\begin{equation}
\lambda_{\#}(p_e(\cdot))= \lambda_N(p_e(\cdot))= \lambda_c(p_e(\cdot))=
\lambda_D(p_e(\cdot)).
\end{equation}
\end{proposition}
\vspace{0.1in}%

\emph{Proof:} The identity $\lambda_{\#}(p_e(\cdot))= \lambda_N(p_e(\cdot))$ is given by
Theorem 6.2 in~\cite{MeRo96}.

We now show $\lambda_{\#}(p_e(\cdot))= \lambda_c(p_e(\cdot))$. The proof method is
similar to the one used for Theorem 3.4 in~\cite{MeRo96}. Suppose
$\lambda>\lambda_{\#}(p_e(\cdot))$. Then for some $\delta>0$,
$\Pr(|W_{\mathbf{0}}'|=\infty)=\delta>0$. For every $h>0$, the box $B(h)=[-h,h]^2$
contains at most a finite number of nodes of
$G(\mathcal{H}_{\lambda,\mathbf{0}},1,p_e(\cdot))$ with probability 1. Thus,
$\Pr(|W_{\mathbf{0}}'\cap B(h)^c|=\infty)=\delta>0$. However, $\{|W_{\mathbf{0}}'\cap
B(h)^c|=\infty\}$ implies $\{|W_{\mathbf{0}}'\cap B(h)^c|>0\}$, so that
$d(W_{\mathbf{0}}')\geq h$. Hence we have $\Pr(d(W_{\mathbf{0}}')\geq h)=\delta>0$. Since
this holds for all $h>0$, we have $\lambda> \lambda_c(p_e(\cdot))$. Therefore,
$\lambda_{\#}(p_e(\cdot))\geq \lambda_c(p_e(\cdot))$.

To show $\lambda_{\#}(p_e(\cdot))\leq \lambda_c(p_e(\cdot))$, note that
$d(W_{\mathbf{0}}')\leq |W_\mathbf{0}'|-1$, where equality is obtained when
$W_{\mathbf{0}}'$ is a chain and the distance between any two adjacent nodes equals 1.
Thus, $\{|W_{\mathbf{0}}'|<\infty\}$ implies $\{d(W_{\mathbf{0}}')<\infty\}$. This proves
$\lambda_{\#}(p_e(\cdot))= \lambda_c(p_e(\cdot))$.

Finally, we show $\lambda_D(p_e(\cdot))=\lambda_N(p_e(\cdot))$. Since
$d(W_{\mathbf{0}}')\leq |W_{\mathbf{0}}'|-1$, $\{E[d(W_{\mathbf{0}}')]=\infty\}$ implies
$\{E[|W_{\mathbf{0}}'|]=\infty\}$. Thus we have $\lambda_D(p_e(\cdot))\geq
\lambda_N(p_e(\cdot))$. On the other hand, if $\lambda>\lambda_N(p_e(\cdot))$, then
$\lambda>\lambda_c(p_e(\cdot))$, i.e., $\Pr(d(W_{\mathbf{0}}')=\infty)>0$. As a
consequence, $E[d(W_{\mathbf{0}}')]=\infty$, which implies $\lambda_N(p_e(\cdot))\geq
\lambda_D(p_e(\cdot))$. Therefore, $\lambda_D(p_e(\cdot))= \lambda_N(p_e(\cdot))$. \qed

Since the four critical densities are identical, in the remainder of this paper, we state
our results with respect to $\lambda_c(p_e(\cdot))$.

It is known that when $\lambda>\lambda_c(p_e(\cdot))$, $G(\mathcal{H}_{\lambda}, 1,
p_e(\cdot))$ is percolated, i.e. with probability 1, there exists a unique infinite
component in $G(\mathcal{H}_{\lambda}, 1)$ consisting of active links and their end
nodes, and when $\lambda<\lambda_c(p_e(\cdot))$, $G(\mathcal{H}_{\lambda}, 1,
p_e(\cdot))$ is not percolated, i.e., with probability 1, there is no infinite component
in $G(\mathcal{H}_{\lambda}, 1)$ consisting of active links and their end
nodes~\cite{MeRo96}.

The following monotonic property for $\lambda_c(p_e(\cdot))$ can be easily proved by
coupling methods.

\vspace{+0.1in}%
\begin{proposition}\label{Proposition-Monotonicity}
Let $\lambda_c(p_e(\cdot))$ and $\lambda_c(p_e'(\cdot))$ be the critical densities for
$G(\mathcal{H}_{\lambda}, 1, p_e(\cdot))$ and $G(\mathcal{H}_{\lambda}, 1, p_e'(\cdot))$,
respectively. Then, if $p_e'(x)\leq p_e(x), \forall x\in (0,1]$, we have
$\lambda_c(p_e(\cdot))\leq\lambda_c(p_e'(\cdot))$.
\end{proposition}
\vspace{+0.1in}%

The following proposition asserts that when the random connection model is in the
subcritical phase, the probability that the origin and a given node are connected decays
exponentially with the distance between them. This is analogous to similar results in
traditional continuum percolation (Theorem 2.4 in~\cite{MeRo96}) and discrete percolation
(Theorem 5.4 in~\cite{Gr99}).

\vspace{+0.1in}%
\begin{proposition}\label{Proposition-Exponential-Decay}
Given $G(\mathcal{H}_{\lambda,\mathbf{0}},1,p_e(\cdot))$ with
$\lambda<\lambda_c(p_e(\cdot))$, let $B(h)=[-h,h]^2$, $h\in \mathbb{R}^+$. Then there
exist constants $c_1,c_2>0$, such that $\Pr(\mathbf{0} \leftrightsquigarrow B(h)^c)\leq
c_1 e^{-c_2h}$, where $\{\mathbf{0} \leftrightsquigarrow B(h)^c\}$ denotes the event that
the origin and some node in $B(h)^c$ are connected, i.e., the origin and some node
outside $B(h)$ are in the same component.
\end{proposition}
\vspace{+0.1in}%

The proof for this proposition is similar to the one for Theorem 2.4 in~\cite{MeRo96}.
For completeness, we give the proof in Appendix A.

\section{Wireless Networks with Dynamic Unreliable Links}

\subsection{Percolation-based Connectivity}

For the random connection model, we assumed that the structure of the graph does not
change with time. Once a link is active, it remains active forever. In wireless networks,
however, the link quality usually varies with time due to shadowing and multi-path
fading. In order to study percolation-based connectivity of wireless networks with
time-varying links, we investigate a more sophisticated model. Formally, given a wireless
network modelled by $G(\mathcal{H}_{\lambda},1)$, we associate a stationary on-off state
process $\{W_{ij}(d_{ij},t);t\geq 0\}$ with each link $(i,j)$, where $d_{ij}$ is the
length of the link, such that $W_{ij}(d_{ij},t)=0$ if link $(i,j)$ is inactive at time
$t$, and $W_{ij}(d_{ij},t)=1$ if link $(i,j)$ is active at time $t$. A similar problem
for discrete lattice has been studied in~\cite{HaPeSt97}. Our model can be viewed as one
of dynamic bond percolation in random geometric graphs.

For such dynamic networks, we will show that there exists a phase transition, and the
critical density for this model is the same as the one for static networks with the
corresponding parameters. To simplify matters, assume that $\{W_{ij}(d_{ij},t)\}$ is
probabilistically identical for all links with the same length.  Use $\{W(d,t)\}$ to
denote the process for a link with length $d$ when no ambiguity arises. Assume that
$\{W(d,t)\}$ is a \emph{Markov} on-off process with i.i.d.~inactive periods $Y_k(d), k
\geq 1$, and i.i.d.~active periods $Z_k(d), k \geq 1$, where $E[Y_k(d) + Z_k(d)] <
\infty$, $\Pr(Z_k(d)> 0) = 1$ and $\Pr(Y_k(d) > 0) = 1$ for $0<d\leq 1$. That is, both
the active and inactive periods are always nonzero. Further assume that $\inf_{0<d\leq
1}\{E[Y_k(d)]\}>0$ and $\sup_{0<d\leq 1}\{E[Y_k(d)]\}<\infty$.

Under the above assumptions, the stationary distribution of $\{W(d,t)\}$ is given
by~\cite{Ro95}
\begin{eqnarray}
\eta_1(d)\triangleq \Pr(W(d,t)=1)=\frac{E[Z_k(d)]}{E[Z_k(d)]+E[Y_k(d)]}, \label{eta-1}\\
\eta_0(d)\triangleq \Pr(W(d,t)=0)=\frac{E[Y_k(d)]}{E[Z_k(d)]+E[Y_k(d)]}, \label{eta-0}
\end{eqnarray}
where $\eta_1(d)$ is the \emph{active ratio} for a link with length $d$.

Let the graph at time $t$ be $G(\mathcal{H}_{\lambda}, 1, W(d,t))$.  That is,
$G(\mathcal{H}_{\lambda}, 1, W(d,t))$ consists of all active links at time $t$, along
with their associated end nodes. The following theorem establishes a phase transition
phenomenon with respect to connectivity in a wireless network with dynamic unreliable
links modelled by $G(\mathcal{H}_{\lambda}, 1, W(d,t))$. It also asserts that the
critical density is the same as the one for the static network $G(\mathcal{H}_{\lambda},
1, \eta_1(d))$, i.e, the network in which each link is active with probability
$\eta_1(d)$.

\vspace{+0.1in}%
\begin{theorem}\label{Theorem-Dynamic-Bond-Percolation}
Let $\lambda_c(\eta_1(d))$ be the critical density for the static model
$G(\mathcal{H}_{\lambda}, 1, \eta_1(d))$. Then $G(\mathcal{H}_{\lambda}, 1, W(d,t))$ is
percolated for all $t\geq 0$ if $\lambda>\lambda_c(\eta_1(d))$, and not percolated at any
$t\geq 0$ if $\lambda<\lambda_c(\eta_1(d))$.
\end{theorem}
\vspace{+0.1in}%

\emph{Proof:} Since $\lambda>\lambda_c(\eta_1(d))$ and $0<\eta_1(d)<1, \forall d\in
(0,1]$, by the monotonic property of $\lambda_c(p_e(\cdot))$ (Proposition
\ref{Proposition-Monotonicity}), we can construct a new model $G(\mathcal{H}_{\lambda},
1, W'(d,t))$ and choose $\epsilon>0$ such that
$\lambda>\lambda_c(\eta_1'(d))\geq\lambda_c(\eta_1(d))$ and $0<\eta_1'(d)<1, \forall d\in
(0,1]$, where $\eta_1'(d)=(1-\epsilon)\eta_1(d)$, for $d\in (0,1]$. As active periods are
always nonzero, we can choose $\delta>0$ such that for any link $(i,j)$,
\[
\Pr(W_{ij}(\delta)=1 | W_{ij}(d,0)=1)>1-\epsilon,
\]
where $W_{ij}(\delta)\triangleq\min_{t\in[0,\delta]} W_{ij}(d,t)$. Then,
\[
\Pr(W_{ij}(\delta)=1)>(1-\epsilon)\eta_1(d)=\eta_1'(d).
\]
Since $\lambda>\lambda_c(\eta_1'(d))$, for any $t\in[0,\delta]$,
$G(\mathcal{H}_{\lambda}, 1, W(d,t))$ is percolated. Repeat this argument for all
intervals $[k\delta,(k+1)\delta]$ with integer $k$. Let $E_k$ be the event that
$G(\mathcal{H}_{\lambda}, 1, W(d,t))$ is percolated for all $t\in[k\delta,(k+1)\delta]$.
Then, we have
\[
\Pr\left(\bigcap_kE_k\right)=1-\Pr\left(\bigcup_kE_k^c\right)\geq 1-\sum_k\Pr(E_k^c)=1.
\]

Similarly, when $\lambda<\lambda_c(\eta_1(d))$, we can construct another model
$G(\mathcal{H}_{\lambda}, 1, W''(d,t))$ and choose $\epsilon>0$ such that
$\lambda<\lambda_c(\eta_1''(d))\leq\lambda_c(\eta_1(d))$ and $0<\eta_1''(d)<1, \forall
d\in (0,1]$, where $\eta_1''(d)=\epsilon(1-\eta_1(d))+\eta_1(d), \forall d\in (0,1]$.
Since inactive periods are always nonzero, we can choose $\delta>0$ such that for any
link $(i,j)$,
\[
\Pr(W_{ij}(\delta)'=0| W_{ij}(d,0)=0)>1-\epsilon,
\]
where $W_{ij}(\delta)'\triangleq \max_{t\in[0,\delta]} W_{ij}(d,t)$. Then,
\[
\Pr(W_{ij}(\delta)'=0)<1-(1-\eta_1(d))(1-\epsilon)=\eta_1''(d).
\]
Since $\lambda<\lambda_c(\eta_1''(d))$, for any $t\in[0,\delta]$,
$G(\mathcal{H}_{\lambda}, 1, W(d,t))$ is not percolated. Repeat this argument for all
intervals $[k\delta,(k+1)\delta]$ with integer $k$, and then proceed in the same way as
before, i.e., using countable additivity. \qed

When the process $\{W(d,t)\}$ is independent of link length $d$, we use $\{W(t)\}$ to
denote the process, and $\eta_1$ and $\eta_0$ to denote its stationary distribution.

\subsection{Information Dissemination in Wireless Networks with Dynamic Unreliable Links}

We have shown that there exists a critical density $\lambda_c(\eta_1(d))$ such that when
$\lambda>\lambda_c(\eta_1(d))$, $G(\mathcal{H}_{\lambda},1,W(d,t))$ is percolated for all
time. If $G(\mathcal{H}_{\lambda},1,W(d,t))$ is percolated, when one node inside the
infinite component of $G(\mathcal{H}_{\lambda},1,W(d,t))$ broadcasts a message to the
whole network, then assuming that there is no propagation delay, all nodes in the infinite component of
$G(\mathcal{H}_{\lambda},1,W(d,t))$ receive this message instantaneously.  The nodes in
the infinite component of $G(\mathcal{H}_{\lambda},1)$ but not in the infinite component
of $G(\mathcal{H}_{\lambda},1,W(d,t))$ cannot receive this message instantaneously.
Nevertheless, as
links switch between the active and inactive states from time to time, those nodes can still
receive the message via multi-hop relaying at some later time.  This remains true even if
$\lambda<\lambda_c(\eta_1(d))$ and $G(\mathcal{H}_{\lambda},1,W(d,t))$ is never
percolated. In this case, when one node inside the infinite component of
$G(\mathcal{H}_{\lambda},1,W(d,t))$ broadcasts a message, due to poor connectivity, only
a small number of nodes can receive this message instantaneously. However, as long as two
nodes $u$ and $v$ are in the infinite component of $G(\mathcal{H}_{\lambda},1)$, the
message can eventually be transmitted from $u$ to $v$ over multi-hop relays. The main
question we address here is the nature of this information dissemination delay.

This problem is similar to the \emph{first passage percolation} problem in
lattices~\cite{Gr99, Ke87}. Related continuum models are considered in~\cite{De03,
DoMaTh04, KoYe07-4}. In \cite{De03}, the author study continuum growth model for a
spreading infection. In \cite{DoMaTh04} and~\cite{KoYe07-4}, the authors consider
wireless sensor networks where each sensor has independent or degree-dependent dynamic
behavior, which can be modelled by an independent or a degree-dependent dynamic site
percolation on random geometric graphs, respectively. The main tool is the Subadditive
Ergodic Theorem~\cite{Li85}. We will use this technique to analyze our problem.

In the following, we will show that in a large-scale wireless network with dynamic
unreliable links, the message delay scales linearly with the Euclidean distance between
the sender and the receiver if the resulting network is in the subcritical phase, and the
delay scales sub-linearly with the distance if the resulting network is in the
supercritical phase.

To begin, we define the delay on a link $(i,j)$ as the amount of time for node $i$ to
deliver a packet to node $j$ over link $(i,j)$. In particular, ignoring propagation
delay, if $(i,j)$ is active when $i$ initiates a transmission, then the delay is zero. If
$(i,j)$ is inactive, the delay is the time from the instant when $i$ initiates
transmission until the instant when $(i,j)$ becomes active. Mathematically, let delay
$T_{ij}(d_{ij})$ be a random variable associated with link $(i,j)$ having length
$d_{ij}$, such that
\begin{equation}\left\{\begin{array}{lll}
\Pr(T_{ij}(d_{ij})=0) & = &\eta_1(d_{ij}), \\
\Pr(T_{ij}(d_{ij})>t) & = &\eta_0(d_{ij}) P_{d_{ij}}(t),
\end{array}\right.
\end{equation}
where $ P_{d_{ij}}(t)=\Pr(W_{ij}(d_{ij},t')=0, \forall t'\in [0,t)|W_{ij}(d_{ij},0)=0)$,
and $(\eta_1(d)$, $\eta_0(d))$ is the stationary distribution of $\{W(d,t)\}$ given by
(\ref{eta-1}) and \eqref{eta-0}.

Let $d(u,v)\triangleq ||\mathbf{X}_u-\mathbf{X}_v||$ and
\begin{equation}\label{eq:T-uv}
T(u,v)=T(\mathbf{X}_u,\mathbf{X}_v)\triangleq\inf_{l(u,v)\in
\mathcal{L}(u,v)}\left\{\sum_{(i,j) \in l(u,v)}T_{ij}(d_{ij})\right\},
\end{equation}
where $l(u,v)$ is a path of adjacent links from node $u$ to node $v$, and
$\mathcal{L}(u,v)$ is the set of all such paths.  Hence, $T(u,v)$ is the message delay on
the path from $u$ to $v$ with the smallest delay.\footnote{\scriptsize Note that the path
with the smallest delay may be different from the shortest path (in terms of number of
links) from node $u$ to node $v$.}

\vspace{+0.1in}%
\begin{theorem}\label{Theorem-Delay}
Given $G(\mathcal{H}_{\lambda},1,W(d,t))$ with $\lambda>\lambda_c$, there exists a
constant $\gamma$ satisfying $\gamma<\infty$ and $\gamma>0$ with probability 1, such that
for any $u,v \in \mathcal{C}(G(\mathcal{H}_{\lambda},1))$, where
$\mathcal{C}(G(\mathcal{H}_{\lambda},1))$ denotes the infinite component of
$G(\mathcal{H}_{\lambda},1)$,
\vspace{+0.1in}%
\begin{itemize}
\item[(i)] if $G(\mathcal{H}_{\lambda},1,W(d,t))$ is in the subcritical phase, i.e.,
$\lambda<\lambda_c(\eta_1(d))$, then for any $\epsilon>0,\delta>0$, there exists
$d_0<\infty$ such that for any $u,v$ with $d(u,v)>d_0$,
\begin{equation}\label{LinearRelation}
\Pr\left(\left|\frac{T(u,v)}{d(u,v)}-\gamma\right|<\epsilon\right)>1-\delta;
\vspace{+0.1in}%
\end{equation}
\item[(ii)] if $G(\mathcal{H}_{\lambda},1,W(d,t))$ is in the supercritical phase, i.e.,
$\lambda>\lambda_c(\eta_1(d))$, then for any $\epsilon>0, \delta>0$, there exists
$d_0<\infty$ such that for any $u,v$ with $d(u,v)>d_0$,
\begin{equation}\label{SubLinearRelation}
\Pr\left(\frac{T(u,v)}{d(u,v)}<\epsilon\right)>1-\delta.
\end{equation}
\end{itemize}
\end{theorem}
\vspace{+0.1in}%

Before proceeding, we introduce some new notation. Let
\begin{eqnarray}
\mathbf{\tilde{X}}_i & \triangleq &\argmin_{\mathbf{X}_j\in
\mathcal{C}(G(\mathcal{H}_{\lambda},1))}\{||\mathbf{X}_j-(i,0)||\},\label{X-i-tilde}\\
T_{l,m} & \triangleq & T(\mathbf{\tilde{X}}_l, \mathbf{\tilde{X}}_m), \mbox{ for }
||\mathbf{\tilde{X}}_l-\mathbf{\tilde{X}}_m||<\infty, 0\leq l\leq m. \label{T-mn}
\end{eqnarray}

The proof for Theorem \ref{Theorem-Delay}-(i) is based on the following lemma:

\vspace{+0.1in}%
\begin{lemma}\label{Lemma-Limit-Convergence}
Let
\begin{equation}\label{eq:gamma}
\gamma\triangleq \lim_{m\rightarrow\infty}\frac{E[T_{0,m}]}{m}.
\end{equation}
Then, $\gamma=\inf_{m\geq 1}\frac{E[T_{0,m}]}{m}$, and
$\lim_{m\rightarrow\infty}\frac{T_{0,m}}{m}=\gamma$ with probability 1.
\end{lemma}
\vspace{+0.1in}%

To show Lemma \ref{Lemma-Limit-Convergence}, we use the following Subadditive Ergodic
Theorem by Liggett~\cite{Li85}.

\vspace{0.1in}%
\begin{theorem}[{Liggett~\cite{Li85}}]\label{Theorem-Subadditive-Ergodic}
Let $\{S_{l,m}\}$ be a collection of random variables indexed by integers $0\leq l<m$.
Suppose $\{S_{l,m}\}$ has the following properties:
\begin{itemize}
\item[(i)] $S_{0,m}\leq S_{0,l}+S_{l,m}, \quad 0\leq l\leq m$; \item[(ii)]
$\{S_{(m-1)k,mk},m\geq 1\}$ is a stationary process for each $k$; \item[(iii)]
$\{S_{l,l+k},k\geq 0\}=\{S_{l+1,l+k+1},k\geq 0\}$ in distribution for each $l$;
\item[(iv)] $E[|S_{0,m}|]<\infty$ for each $m$.
\end{itemize}
Then
\begin{itemize}
\item[(a)] $\alpha\triangleq \lim_{m\rightarrow\infty}\frac{E[S_{0,m}]}{m} = \inf_{m\geq
1}\frac{E[S_{0,m}]}{m}$; $S\triangleq \lim_{m\rightarrow\infty}\frac{S_{0,m}}{m}$ exists
with probability 1 and $E[S]=\alpha$.
\end{itemize}
Furthermore, if
\begin{itemize}
\item[(v)] the stationary process in (ii) is ergodic,
\end{itemize}
then
\begin{itemize}
\item[(b)] $S=\alpha$ with probability 1.
\end{itemize}
\end{theorem}
\vspace{0.1in}%

To show Lemma \ref{Lemma-Limit-Convergence}, we need to verify that the sequence
$\{T_{l,m},l\leq m\}$ satisfies conditions (i)--(v) of Theorem
\ref{Theorem-Subadditive-Ergodic}. It is easy to see that (i) is satisfied, since
$T_{0,m}$ is the delay of the path with the smallest delay from $\mathbf{\tilde{X}}_0$ to
$\mathbf{\tilde{X}}_m$ and $T_{0,l}+T_{l,m}$ is the delay on a particular path from
$\mathbf{\tilde{X}}_0$ to $\mathbf{\tilde{X}}_l$ (it has the smallest delay from
$\mathbf{\tilde{X}}_0$ to $\mathbf{\tilde{X}}_l$, and from $\mathbf{\tilde{X}}_l$ to
$\mathbf{\tilde{X}}_m$). Furthermore, because all nodes are distributed according to a
homogeneous Poisson point process, the geometric structure is stationary and hence (ii)
and (iii) are guaranteed. We need only to show conditions (iv) and (v) also hold for
$\{T_{l,m},l\leq m\}$. To accomplish this, we first show property (iv) holds for
$\{T_{l,m},l\leq m\}$.

\vspace{+0.1in}%
\begin{lemma}\label{Lemma-Finite-Distance}
Let $r_0=||\tilde{\mathbf{X}}_0-(0,0)||$, then $r_0<\infty$ with probability 1.
\end{lemma}
\vspace{+0.1in}%

\emph{Proof:} We consider a mapping between $G(\mathcal{H}_{\lambda},1)$ and a square
lattice $\mathcal{L}=d\cdot\mathbb{Z}^2$, where $d$ is the edge length. The vertices of
$\mathcal{L}$ are located at $(d\times i, d\times j)$ where $(i,j)\in \mathbb{Z}^2$. For
each horizontal edge $a$, let the two end vertices be $(d\times a_x, d\times a_y)$ and
$(d\times a_x+d, d\times a_y)$.

\begin{figure}[t!]
\centerline{ \subfigure[Good Rectangle]{
\includegraphics[width=3in]{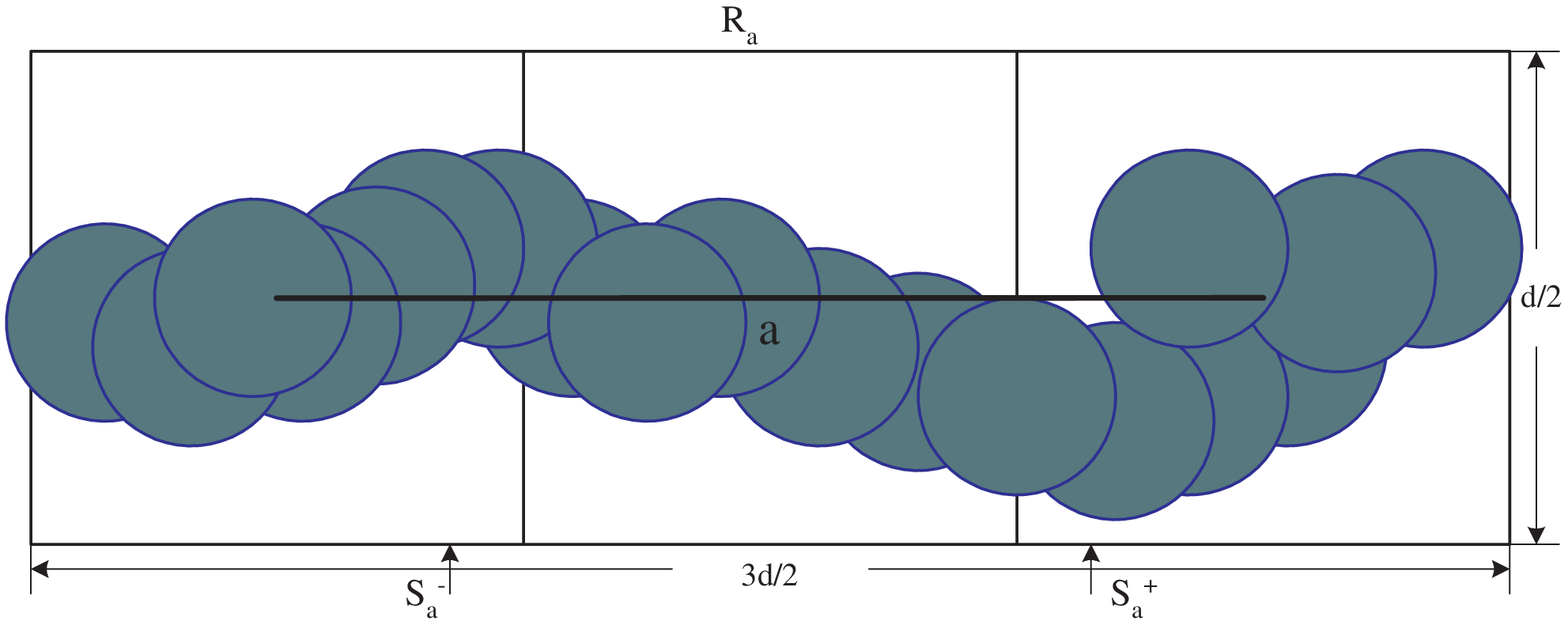}}\hfil
\subfigure[Open Rectangle]{
\includegraphics[width=3in]{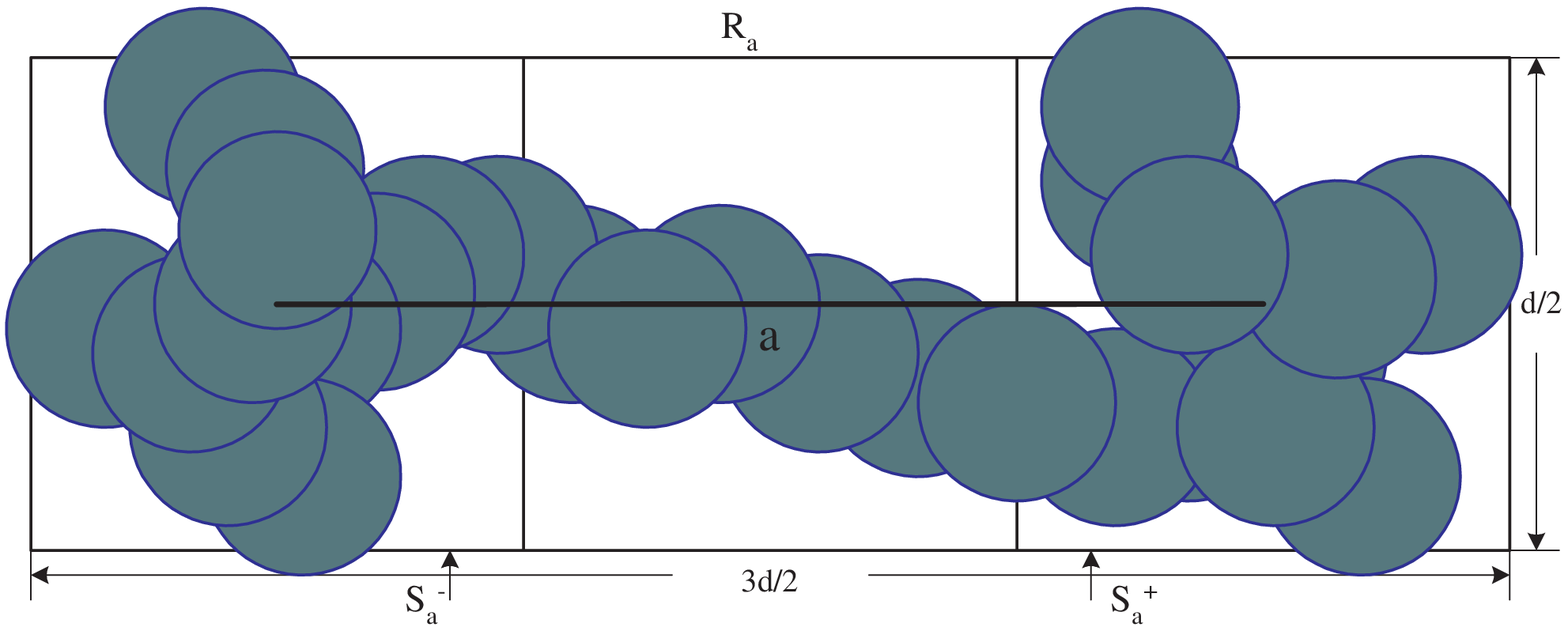}}}
\caption{Examples of good and open rectangles (edges)}\label{fig:GoodOpenRectangle}
\end{figure}

For edge $a$ in $\mathcal{L}$, define event $A_a(d)$ as the set of outcomes for which the
following condition holds: the rectangle $R_a=[a_xd-\frac{d}{4},a_xd+\frac{5d}{4}]\times
[a_yd-\frac{d}{4},a_yd+\frac{d}{4}]$ is crossed\footnote{\scriptsize Here, a rectangle
$R=[x_1,x_2]\times[y_1,y_2]$ being crossed from left to right by a connected component in
$G(\mathcal{H}_\lambda,1)$ means that there exists a sequence of nodes
$v_1,v_2,...,v_m\in G(\mathcal{H}_\lambda,1)$ contained in $R$, with
$||\mathbf{x}_{v_i}-\mathbf{x}_{v_{i+1}}||\leq 1, i=1,...,m-1$, and $0<x(v_1)-x_1<1,
0<x_2-x(v_m)<1$, where $x(v_1)$ and $x(v_m)$ are the $x$-coordinates of nodes $v_1$ and
$v_m$, respectively.  A rectangle being crossed from top to bottom is defined
analogously.} from left to right by a connected component in $G(\mathcal{H}_\lambda,1)$.
If $A_a(d)$ occurs, we say that rectangle $R_a$ is a {\em good} rectangle, and edge $a$
is a {\em good} edge. Let
\[
p_g(d)\triangleq\Pr(A_a(d)).
\]
Define $A_a(d)$ similarly for all vertical edges by rotating the rectangle by
$90^{\circ}$. An example of a good rectangle and a good edge is illustrated in
Figure~\ref{fig:GoodOpenRectangle}-(a).

Further define event $B_a(d)$ for edge $a$ in $\mathcal{L}$ as the set of outcomes for
which both of the following hold: (i) $A_a(d)$ occurs; (ii) the left square
$S_a^-=[a_xd-\frac{d}{4},a_xd+\frac{d}{4}]\times [a_yd-\frac{d}{4},a_yd+\frac{d}{4}]$ and
the right square $S_a^+=[a_xd+\frac{3d}{4},a_xd+\frac{5d}{4}]\times
[a_yd-\frac{d}{4},a_yd+\frac{d}{4}]$ are both crossed from top to bottom by connected
components in $G_1(\mathcal{H}_\lambda,1)$.

If $B_a(d)$ occurs, we say that rectangle $R_a$ is an {\em open} rectangle, and edge $a$
is an {\em open} edge. Let
\[
p_o(d)\triangleq\Pr(B_a(d)).
\]
Define $B_a(d)$ similarly for all vertical edges by rotating the rectangle by
$90^{\circ}$. Examples of an open rectangle and an open edge are illustrated in
Figure~\ref{fig:GoodOpenRectangle}-(b).

Suppose edges $b$ and $c$ are vertically adjacent to edge $a$, then it is clear that if
events $A_a(d)$, $A_b(d)$ and $A_c(d)$ all occur, then event $B_a(d)$ occurs. Moreover,
since events $A_a(d)$, $A_b(d)$ and $A_c(d)$ are increasing events\footnote{\scriptsize
An event $A$ is called increasing if $I_A(G)\leq I_A(G')$ whenever graph $G$ is a
subgraph of $G'$, where $I_A$ is the indicator function of $A$. An event $A$ is called
decreasing if $A^{c}$ is increasing. For details, please see~\cite{Gr99, MeRo96, Pe03}.},
by the FKG inequality~\cite{Gr99, MeRo96, Pe03},
\begin{eqnarray*}
p_o(d) & = & \Pr(B_a(d))\\
&\geq&\Pr(A_a(d)\cap A_b(d)\cap A_c(d))\\
&\geq&\Pr(A_a(d))\Pr(A_b(d))\Pr(A_c(d))\\
& = &(p_g(d))^3.
\end{eqnarray*}

According to Corollary 4.1 in~\cite{MeRo96}, the probability $p_g(d)$ converges to 1 as
$d\rightarrow \infty$ when $G(\mathcal{H}_{\lambda},1)$ is in the supercritical phase. In
this case, $(p_g(d))^3$ converges to 1 as $d\rightarrow \infty$ as well. Hence, $p_o(d)$
converges to 1 as $d\rightarrow \infty$ when $G(\mathcal{H}_{\lambda},1)$ is in the
supercritical phase.

Note that in our model, the events $\{B_a(d)\}$ are not independent in general. However,
if two edges $a$ and $b$ are not adjacent, i.e., they do not share any common end
vertices, then $B_a(d)$ and $B_b(d)$ are independent. Furthermore, when edges $a$ and $b$
are adjacent, $B_a(d)$ and $B_b(d)$ are increasing events and thus positively
correlated~\footnote{Positive correlation means $\Pr(B_a(d)|B_b(d))>\Pr(B_a(d))$.}.
Consequently, our model is a 1-dependent bond percolation model. It is known that there
exists $p^{\mbox{\small{bond}}}_{\mbox{\small{1-dep}}}<1$ such that any 1-dependent model
with $p>p^{\mbox{\small{bond}}}_{\mbox{\small{1-dep}}}$ is percolated, where $p$ is the
probability of an edge being open~\cite{LiScSt97}.

Now define
\begin{equation}
d_0\triangleq\inf\left\{d'>1:p_o(d')>\max\left\{\frac{8}{9},
p^{\mbox{\small{bond}}}_{\mbox{\small{1-dep}}}\right\}\right\},
\end{equation}
and choose the edge length of $\mathcal{L}$ to be $d>d_0$. Then there is an infinite
cluster consisting of open edges and their end vertices in $\mathcal{L}$. Denote this
infinite cluster by $\mathcal{C}(\mathcal{L})$.

\begin{figure}[t]
\centering
\includegraphics[width=2.5in]{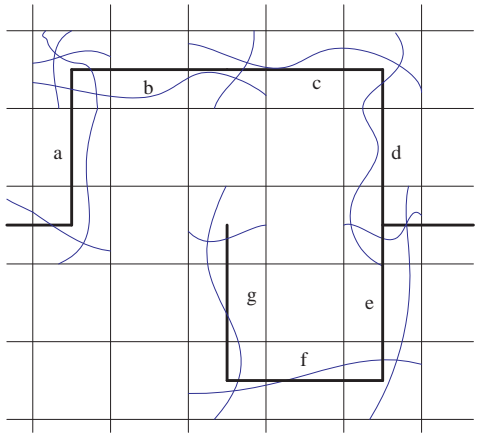}
\caption{A path of open edges in $\mathcal{L}$ implies a path of connected nodes in
$G(\mathcal{H}_{\lambda},1)$}\label{fig:OpenPath}
\end{figure}

From Figure~\ref{fig:OpenPath}, it is easy to see that all the nodes along the crossings
in $R_a$ and all the nodes along the crossings in $R_b$ for any $a, b\in
\mathcal{C}(\mathcal{L})$ are connected. Since the infinite component of
$G(\mathcal{H}_{\lambda},1)$ is unique, all the nodes along the crossings in $R_a$ for
each $a\in \mathcal{C}(\mathcal{L})$ must belong to
$\mathcal{C}(G(\mathcal{H}_{\lambda},1))$.

\begin{figure}[t!]
\centerline{ \subfigure[]{
\includegraphics[width=3in]{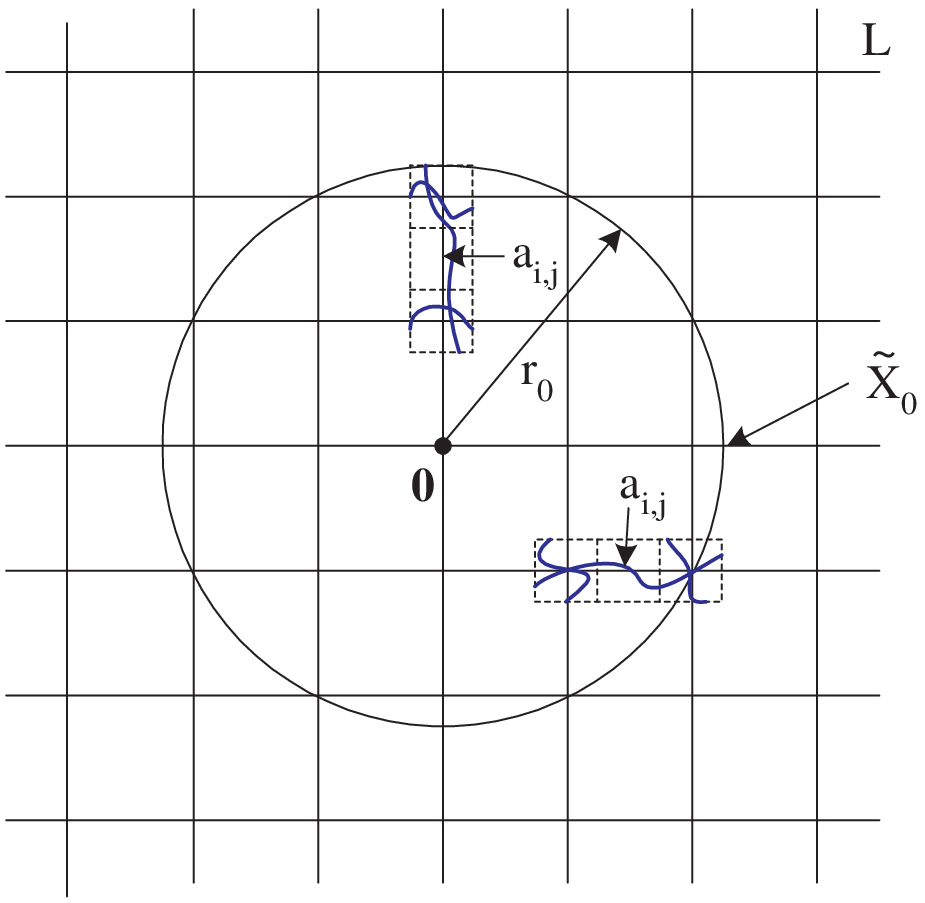}}\hfil
\subfigure[]{
\includegraphics[width=3in]{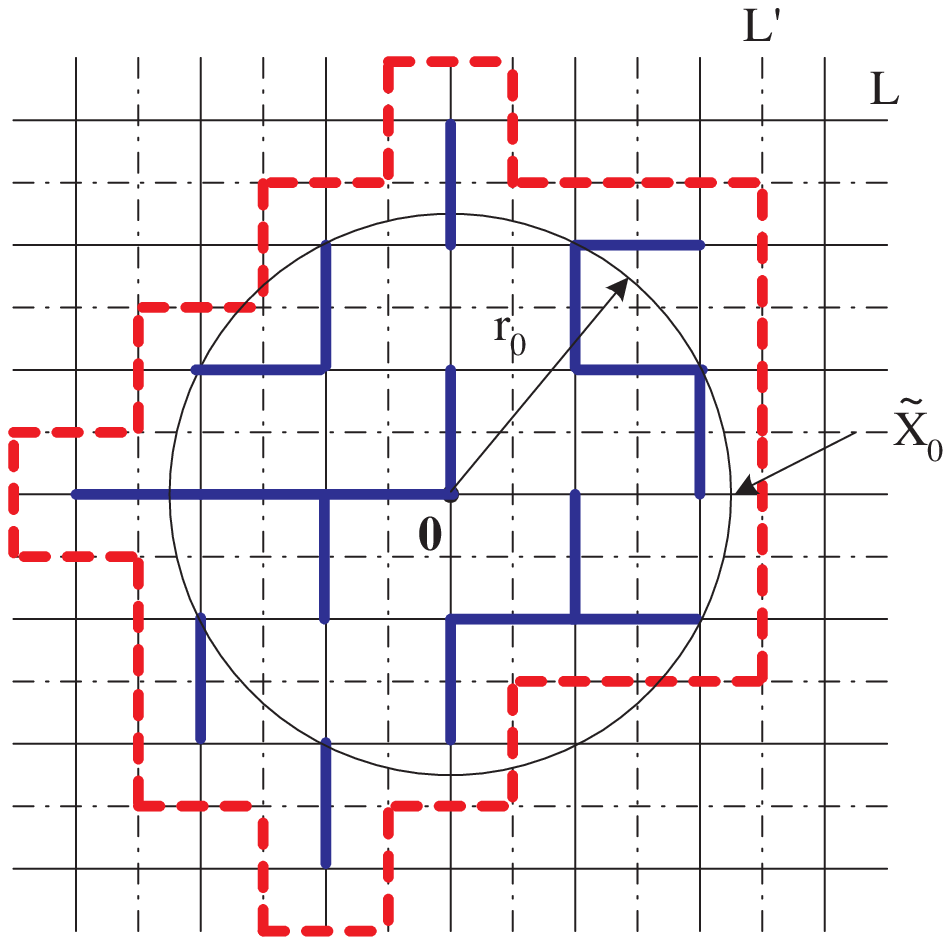}}}
\caption{(a) Two possibilities for $a_{i,j}$ in $\mathcal{L}$. (b) There exists a closed
circuit in $\mathcal{L}$ containing all edges of $\mathcal{L}$ that are strictly inside
$\mathcal{A}(\mathbf{0},r_0)$}\label{fig:Circle}
\end{figure}

By definition, no node of $G(\mathcal{H}_{\lambda},1)$ strictly inside
$\mathcal{A}(\mathbf{0},r_0)$ belongs to $\mathcal{C}(G(\mathcal{H}_{\lambda},1))$. This
implies that no edge of $\mathcal{L}$ strictly inside $\mathcal{A}(\mathbf{0},r_0)$
belongs to $\mathcal{C}(\mathcal{L})$. To see this, suppose edge $a_{i,j}$ of
$\mathcal{L}$ is strictly inside $\mathcal{A}(\mathbf{0},r_0)$ and belongs to
$\mathcal{C}(\mathcal{L})$. The nodes along the crossings in $R_{a_{i,j}}$ belong to
$\mathcal{C}(G(\mathcal{H}_{\lambda},1))$. As shown in Figure~\ref{fig:Circle}-(a), when
$d>1$ and $r_0\gg1$, no matter what direction the edge $a_{i,j}$ has, there are some
nodes along the crossings in $R_{a_{i,j}}$ (therefore belonging to
$\mathcal{C}(G(\mathcal{H}_{\lambda},1))$) which are strictly inside
$\mathcal{A}(\mathbf{0},r_0)$. These nodes then have strictly smaller distance to
$\mathbf{0}$ than node $\mathbf{\tilde{X}}_0$. This contradiction ensures that no edge of
$\mathcal{L}$ strictly inside $\mathcal{A}(\mathbf{0},r_0)$ belongs to
$\mathcal{C}(\mathcal{L})$.

Consider the {\em dual lattice} $\mathcal{L}'$ of $\mathcal{L}$. The construction of
$\mathcal{L}'$ is as follows: let each vertex of $\mathcal{L}'$ be located at the center
of a square of $\mathcal{L}$. Let each edge of $\mathcal{L}'$ be open if and only if it
crosses an open edge of $\mathcal{L}$, and closed otherwise. It is clear that each edge
in $\mathcal{L}'$ is open also with probability $p_o(d)$. Let
\[
q=1-p_o(d)<\frac{1}{9}.
\]

Choose $2m$ edges in $\mathcal{L}'$. Since the states (open or closed) of any set of
non-adjacent edges are independent, we can choose $m$ edges among the $2m$ edges such
that their states are independent. As a result,
\[
\Pr(\mbox{all the $2m$ edges are closed})\leq q^m.
\]

Now a key observation is that if no edge of $\mathcal{L}$ strictly inside
$\mathcal{A}(\mathbf{0},r_0)$ belongs to $\mathcal{C}(\mathcal{L})$, for which the event
is denoted by $E_{\mathcal{L}}$, then there must exist a closed circuit in $\mathcal{L}'$
(a circuit consisting of closed edges) containing all edges of $\mathcal{L}$ strictly
inside $\mathcal{A}(\mathbf{0},r_0)$, for which the event is denoted by
$E_{\mathcal{L}'}$, and vice versa~\cite{Gr99}. This is demonstrated in
Figure~\ref{fig:Circle}-(b). Hence
\[
\Pr(E_{\mathcal{L}})=1 \Longleftrightarrow \Pr(E_{\mathcal{L}'})=1.
\]

Any closed circuit in $\mathcal{L}'$ containing all edges of $\mathcal{L}$ strictly
inside $\mathcal{A}(\mathbf{0},r_0)$ has length greater than or equal to $2l$, where
$l\triangleq 2\lfloor \frac{r_0}{d}\rfloor$. Thus we have
\[
\Pr(E_{\mathcal{L}'}) = \sum_{m=l}^{\infty}\Pr(\exists \mathcal{O}_c(2m)) \leq
\sum_{m=l}^{\infty} \gamma(2m)q^m,
\]
where $\mathcal{O}_c(2m)$ is a closed circuit having length $2m$ in $\mathcal{L}'$
containing all edges of $\mathcal{L}$ strictly inside $\mathcal{A}(\mathbf{0},r_0)$, and
$\gamma(2m)$ is the number of such circuits. By
Proposition~\ref{Lemma-Closed-Circuit-Number} in Appendix B, we have
$\gamma(2m)=\frac{4}{27}(m-1)3^{2m}$ so that
\begin{eqnarray}
\sum_{m=l}^{\infty} \gamma(2m)q^m &\leq & \sum_{m=l}^{\infty} \frac{4}{27}(m-1)(9q)^m\nonumber\\
&= & \frac{4}{27}\left[\sum_{m=l}^{\infty} m(9q)^m-\sum_{m=l}^{\infty}(9q)^m\right]\nonumber\\
& = &\frac{4[l-1-(l-2)9q]}{27(1-9q)^2}(9q)^l.
\end{eqnarray}
Since $q<\frac{1}{9}$, we have $\Pr(E_{\mathcal{L}'})\rightarrow 0$ as $l=2\lfloor
\frac{r_0}{d}\rfloor\rightarrow \infty$. That is, as $r_0$ goes to infinity, with
probability 1, there is some edge of $\mathcal{L}$ strictly inside
$\mathcal{A}(\mathbf{0},r_0)$ belonging to $\mathcal{C}(\mathcal{L})$. Hence, with
probability 1, there is some node of $G(\mathcal{H}_{\lambda},1)$ strictly inside
$\mathcal{A}(\mathbf{0},r_0)$ belonging to $\mathcal{C}(G(\mathcal{H}_{\lambda},1))$.
This contradiction implies that $r_0$ is finite with probability 1.\qed

Let $r_m=||\tilde{\mathbf{X}}_m-(m,0)||$, by Lemma~\ref{Lemma-Finite-Distance} and
stationarity, we have $r_m<\infty$ with probability 1, for any $m$.

\vspace{+0.1in}%
\begin{lemma}\label{Lemma-Finite-Hops}
Let $L(\mathbf{\tilde{X}}_0,\mathbf{\tilde{X}}_m)$ be the shortest path (in terms of the
number of links) from $\mathbf{\tilde{X}}_0$ to $\mathbf{\tilde{X}}_m$, and let
$|L(\mathbf{\tilde{X}}_0,\mathbf{\tilde{X}}_m)|$ denote the number of links on such a
path. If $||\mathbf{\tilde{X}}_0-\mathbf{\tilde{X}}_m||<\infty$, then
$|L(\mathbf{\tilde{X}}_0,\mathbf{\tilde{X}}_m)|<\infty$, and $E[T_{0,m}^L]<\infty$, where
$T_{0,m}^L$ denotes the delay on path $L(\mathbf{\tilde{X}}_0,\mathbf{\tilde{X}}_m)$.
\end{lemma}
\vspace{+0.1in}%

\emph{Proof:} We use the same mapping as the one for the proof of
Lemma~\ref{Lemma-Finite-Distance}. For any given $\sqrt[4]{\frac{8}{9}}<\delta<1$, define
\begin{equation}
d_{\delta}=\max\{\inf\{d':p_g(d')\geq \delta\},
||\mathbf{\tilde{X}}_0-\mathbf{\tilde{X}}_m||\}.
\end{equation}
Then, for any $d>d_{\delta}$, we have $p_g(d)\geq \delta$.

\begin{figure}[t]
\centering
\includegraphics[width=2.5in]{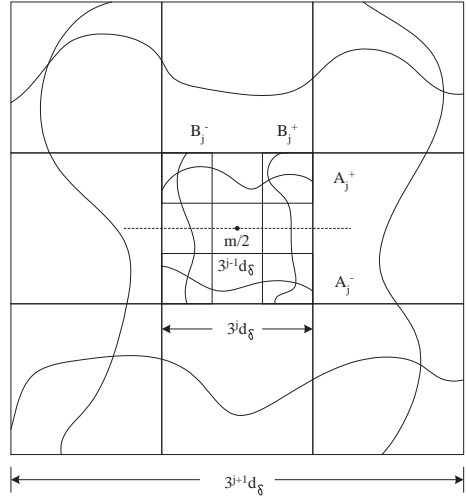}
\caption{Square annuli}\label{fig:SquareAnnulus}
\end{figure}

Now, consider a fractal structure as shown in Figure~\ref{fig:SquareAnnulus}: first a
square $S(d_{\delta})$ is constructed with edge length $d_{\delta}$ centered at
$\frac{\mathbf{\tilde{X}}_0+\mathbf{\tilde{X}}_m}{2}$. Then, a second square
$S(3d_{\delta})$ is constructed with edge length $3d_{\delta}$ also centered at
$\frac{\mathbf{\tilde{X}}_0+\mathbf{\tilde{X}}_m}{2}$. The construction proceeds in the
same manner, i.e., at step $j$, a square $S(3^{j-1}d_{\delta})$ is constructed with edge
length $3^{j-1}d_{\delta}$ centered at
$\frac{\mathbf{\tilde{X}}_0+\mathbf{\tilde{X}}_m}{2}$. Thus, we have the initial square
and a sequence of square annuli that do not overlap.

Denote the square annulus with inside edge length $3^{j-1}d_{\delta}$ ($j \geq 2$) and
outside edge length $3^jd_{\delta}$ by $D(3^jd_{\delta})$. Let $A_j^+$ be the event that
the upper horizontal rectangle of $D(3^jd_{\delta})$---
$[\frac{m}{2}-\frac{3^j}{2}d_{\delta},\frac{m}{2}+\frac{3^j}{2}d_{\delta}]
\times[\frac{3^{j-1}}{2}d_{\delta}, \frac{3^j}{2}d_{\delta}]$ is good, i.e., it is
crossed by a connected component in $G(\mathcal{H}_{\lambda},1)$ from left to right.
Since the length of the corresponding lattice edge of the upper horizontal rectangle of
$D(3^jd_{\delta})$ is $2\cdot3^{j-1}d_{\delta}>d_{\delta}$, we have
$\mbox{Pr}\{A_j^+\}\geq \delta$. Similarly define $A_j^-$, $B_j^+$ and $B_j^-$ to be the
events that the lower, right and left rectangles are good, respectively. Then
$\mbox{Pr}\{A_j^-\}\geq \delta$, $\mbox{Pr}\{B_j^+\}\geq \delta$ and
$\mbox{Pr}\{B_j^-\}\geq \delta, \forall j\geq 1$.

Let $E_j$ be the event that there exists a circuit of connected nodes in
$G(\mathcal{H}_{\lambda},1)$ within $D(3^jd_{\delta})$. Once $A_j^+, A_j^-, B_j^+$ and
$B_j^-$ all occur, $E_j$ must also occur. Although $A_j^+, A_j^-, B_j^+$ and $B_j^-$ are
not independent, they are increasing events. By the FKG inequality, we have
\begin{eqnarray}
\Pr(E_j) & \geq & \Pr(A_j^+\cap A_j^-\cap B_j^+\cap B_j^-)\nonumber\\
& \geq & \Pr(A_j^+)\Pr(A_j^-) \Pr(B_j^+)\Pr(B_j^-)\nonumber\\
& \geq & \delta^4.
\end{eqnarray}

When $E_j$ occurs, $\mathbf{\tilde{X}}_0$ and $\mathbf{\tilde{X}}_m$ are contained in
$S(3^{j-1}d_{\delta})$ and there is a circuit of connected nodes in
$G(\mathcal{H}_{\lambda},1)$ contained in the square annulus $D(3^jd_{\delta})$. If the
shortest path between $\mathbf{\tilde{X}}_0$ and $\mathbf{\tilde{X}}_m$,
$L(\mathbf{\tilde{X}}_0, \mathbf{\tilde{X}}_m)$, were to go outside $S(3^jd_{\delta})$,
it would intersect the closed circuit contained by $D(3^jd_{\delta})$ and we could
construct a shorter path from $\mathbf{\tilde{X}}_0$ to $\mathbf{\tilde{X}}_m$. This
implies that $L(\mathbf{\tilde{X}}_0, \mathbf{\tilde{X}}_m)$ must be contained in
$S(3^jd_{\delta})$.

Suppose $u$, $v$ and $w$ are three consecutive nodes along $L(\mathbf{\tilde{X}}_0,
\mathbf{\tilde{X}}_m)$. Then $||\mathbf{X}_u-\mathbf{X}_w||> 1$, since otherwise $v$
would not belong to the shortest path. Hence, if we draw circles with radius
$\frac{1}{2}$, centered at $\mathbf{X}_u$ and $\mathbf{X}_w$, respectively, then the two
circles do not overlap. Consequently, if the length of $L(\mathbf{\tilde{X}}_0,
\mathbf{\tilde{X}}_m)$ is $|L|\triangleq|L(\mathbf{\tilde{X}}_0, \mathbf{\tilde{X}}_m)|$,
then we must be able to draw at least $\lceil\frac{|L|}{2}\rceil$ circles with radius
$\frac{1}{2}$ centered at alternating nodes along $L(\mathbf{\tilde{X}}_0,
\mathbf{\tilde{X}}_m)$. All of these circles are contained in the square with edge length
$3^jd_{\delta}+1$. Such a square contains at most $\lceil
(3^jd_{\delta}+1)^2/[\pi(\frac{1}{2})^2] \rceil$ non-overlapping circles with radius
$\frac{1}{2}$. Therefore, $|L|\leq 2 \lceil 4(3^jd_{\delta}+1)^2/\pi\rceil<\infty$.

Now if $|L|> 2 \lceil 4(3^jd_{\delta}+1)^2/\pi\rceil$, then $|L|> 2 \lceil
4(3^id_{\delta}+1)^2/\pi\rceil$ for all $i=1,2,...,j$.  By the above argument, none of
the events $E_1, E_2,...E_j$ can occur. Thus
\[
\Pr\left(|L|> 2 \left\lceil \frac{4}{\pi}(3^jd_{\delta}+1)^2\right\rceil\right)\leq
\prod_{i=1}^j\Pr(E_i^c)\leq (1-\delta^4)^j.
\]

Let $M=2 \left\lceil \frac{4}{\pi}(3d_{\delta}+1)^2\right\rceil$, then we have
\begin{eqnarray}
E[|L|]&=&\sum_{k=0}^{\infty}\Pr(|L|> k)\nonumber\\
&=&\sum_{k=0}^{M}\Pr(|L|> k)+\sum_{k=M+1}^{\infty}\Pr(|L|> k)\nonumber\\
&\leq& M+ \sum_{j=1}^{\infty}\left\lceil
\frac{4}{\pi}(3^{j+1}d_{\delta}+1)^2\right\rceil\Pr\left(|L|> \left\lceil
\frac{4}{\pi}(3^jd_{\delta}+1)^2\right\rceil\right)\nonumber\\
&\leq& M+ \sum_{j=1}^{\infty}\left(
\frac{4}{\pi}(3^{j+1}d_{\delta}+1)^2+1\right) (1-\delta^4)^j\nonumber\\
&=& M+ \sum_{j=1}^{\infty}\left(
\frac{4}{\pi}(9\cdot9^jd^2_{\delta}+6\cdot3^jd_{\delta}+1)+1\right) (1-\delta^4)^j\nonumber\\
&=& M+ \frac{36d^2_{\delta}}{\pi}\sum_{j=1}^{\infty}9^j(1-\delta^4)^j
+\frac{24d_{\delta}}{\pi}\sum_{j=1}^{\infty}3^j(1-\delta^4)^j
+\left(\frac{4}{\pi}+1\right)\sum_{j=1}^{\infty}(1-\delta^4)^j.
\end{eqnarray}
When $\delta>\sqrt[4]{\frac{8}{9}}$, we have $(1-\delta^4)^j<9^{-j}$. Thus,
$E[|L|]<\infty$.

Let $\Lambda_{W(d,t)}\triangleq \sup_{0<d\leq 1}\{\eta_0(d)E[Y_k(d)]\}<\infty$, then
\begin{equation}
E[T_{0,m}^{L}||L|]=\sum_{i=1}^{|L|}\eta_0^{(i)}(d)E[Y_k^{(i)}(d)]\leq
|L|\Lambda_{W(d,t)},
\end{equation}
where $\eta_0^{(i)}(d)$ and $E[Y_k^{(i)}(d)]$ are the stationary probability of the
inactive state, and the expected inactive period of the $i$-th link with length $d$ on
$L(\mathbf{\tilde{X}}_0, \mathbf{\tilde{X}}_m)$, respectively. Hence
\begin{equation}
E[T_{0,m}^{L}]=E[E[T_{0,m}^{L}||L|]]\leq E[|L|]\Lambda_{W(d,t)}<\infty.
\end{equation}
\qed

To show property (v), we show $\{T_{(m-1)j,mj}, m\geq 1\}$ is strong
mixing.\footnote{\scriptsize A measure preserving transformation $H$ on $(\Omega,
\mathcal{F},P)$ is called strong mixing if for all measurable sets $A$ and $B$,
$\lim_{m\rightarrow\infty}|P(A\cap H^{-m}B)-P(A)P(B)|=0$. A sequence $\{X_n,n\geq 0\}$ is
called strong mixing if the shift on sequence space is strong (weak) mixing. Every strong
mixing system is ergodic~\cite{Du96}.}

\vspace{+0.1in}%
\begin{lemma}\label{Lemma-Strongly-Mixing}
The sequence $\{T_{(m-1)k,mk}, m\geq 1\}$ is strong mixing, so that it is ergodic.
\end{lemma}
\vspace{+0.1in}%

\emph{Proof:} From the proof of Lemma~\ref{Lemma-Finite-Distance}, we have $\Pr(E_j)\geq
\delta^4$ for all $j=1,2,...$. Summing over $j$ yields
\begin{equation}
\sum_{j=1}^{\infty}\Pr(E_j)\geq \sum_{j=1}^{\infty}\delta^4=\infty.
\end{equation}
Since ${E_j}$ are independent events,  by the Borel-Cantelli Lemma, with probability 1,
there exists $j'<\infty$ such that $E_{j'}$ occurs.

We now construct squares $A_1$ and $A_2$ centered at
$\frac{\mathbf{\tilde{X}}_{(m-1)j}+\mathbf{\tilde{X}}_{mj}}{2}$ and
$\frac{\mathbf{\tilde{X}}_{(m+k-1)j}+\mathbf{\tilde{X}}_{(m+k)j}}{2}$ with edge length
$3^{j'}d_{\delta}$ and $3^{j''}d_{\delta}$ respectively, such that the path with the
smallest delay from $\mathbf{\tilde{X}}_{(m-1)j}$ to $\mathbf{\tilde{X}}_{mj}$, and the
path with the smallest delay from $\mathbf{\tilde{X}}_{(m+k-1)j}$ to
$\mathbf{\tilde{X}}_{(m+k)j}$ are contained in $A_1$ and $A_2$, respectively. Let $E$ be
the event that $j'<\infty$ and $j''<\infty$. Then $\Pr(E)=1$.

\begin{figure}[t]
\centering
\includegraphics[width=3.5in]{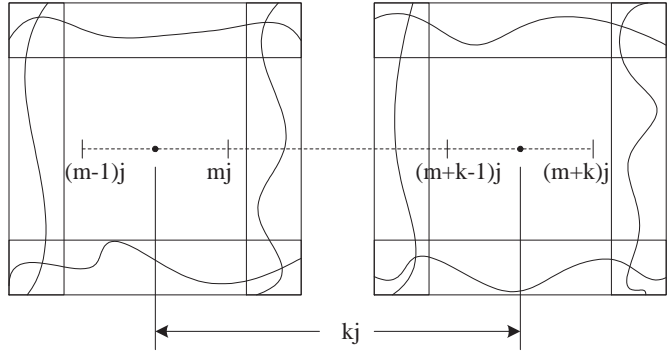}
\caption{As $k\rightarrow \infty$, the paths inside $A_1$ and $A_2$ do not share any
common nodes. Hence $T_{(m-1)j,mj}$ and $T_{(m+k-1)j,(m+k)j}$ are independent of each
other as $k\rightarrow \infty$.}\label{fig:TwoAnnuli}
\end{figure}

When finite $j'$ and $j''$ exist, due to stationarity,  $j'$ and $j''$ are independent of
$k$. Hence, as $k\rightarrow\infty$, $A_1$ and $A_2$ become non-overlapping so that the
paths inside $A_1$ and $A_2$ do not share any common nodes of
$G(\mathcal{H}_{\lambda},1)$. Hence $T_{(m-1)j,mj}$ and $T_{(m+k-1)j,(m+k)j}$ are
independent of each other as $k\rightarrow \infty$. This is illustrated in
Figure~\ref{fig:TwoAnnuli}.

Therefore
\begin{eqnarray}
&&\lim_{k\rightarrow\infty}\Pr(\{T_{(m-1)j,mj}<t\}\cap \{T_{(m+k-1)j,(m+k)j}<t'\})\nonumber\\
& =&\lim_{k\rightarrow\infty}\Pr(\{T_{(m-1)j,mj}<t\}\cap
\{T_{(m+k-1)j,(m+k)j}<t'\}|E)\Pr(E)\nonumber\\
&&+\lim_{k\rightarrow\infty}\Pr(\{T_{(m-1)j,mj}<t\}\cap \{T_{(m+k-1)j,(m+k)j}<t'\}|E^c)\Pr(E^c)\nonumber\\
&=&\Pr(T_{(m-1)j,mj}<t|E)\Pr(T_{(m-1)j,mj}<t'|E)\nonumber\\
&=&\Pr(T_{(m-1)j,mj}<t)\Pr(T_{(m-1)j,mj}<t'),
\end{eqnarray}
This implies that sequence $\{T_{(m-1)k,mk}, m\geq 1\}$ is strong mixing, so that it is
ergodic.\qed

Now, we present the proof for Lemma \ref{Lemma-Limit-Convergence}.

\emph{Proof of Lemma \ref{Lemma-Limit-Convergence}:} Conditions (i)--(iii) of
Theorem~\ref{Theorem-Subadditive-Ergodic} have been verified. The validation of (iv) is
provided by Lemma~\ref{Lemma-Finite-Hops}. Let
$L(\mathbf{\tilde{X}}_0,\mathbf{\tilde{X}}_m)$ be the shortest path from
$\mathbf{\tilde{X}}_0$ to $\mathbf{\tilde{X}}_m$. Since
$L(\mathbf{\tilde{X}}_0,\mathbf{\tilde{X}}_m)$ is a particular path, we have $T_{0,m}\leq
T_{0,m}^L$ so that $E[T_{0,m}]\leq E[T_{0,m}^L]$, where $T_{0,m}^L$ denotes the delay on
path $L(\mathbf{\tilde{X}}_0,\mathbf{\tilde{X}}_m)$. By Lemma~\ref{Lemma-Finite-Hops}, we
have $E[T_{0,m}^{L}]<\infty$ and therefore $E[T_{0,m}]<\infty$. Furthermore, due to Lemma
\ref{Lemma-Strongly-Mixing}, $\{T_{(m-1)k,mk}, m\geq 1\}$ is ergodic, thus the results
(a) and (b) of Theorem \ref{Theorem-Subadditive-Ergodic} hold.\qed

\emph{Remark:} Using the proof for condition (iv) of
Theorem~\ref{Theorem-Subadditive-Ergodic}, we can show that for any two nodes $u$ and $v$
in the infinite component of $G(\mathcal{H}_{\lambda},1)$ which are within finite
Euclidean distance of each other, i.e., $u,v\in \mathcal{C}(G(\mathcal{H}_{\lambda},1))$
with $d(u,v)<\infty$, $E[T(u,v)]<\infty$.

The following lemma asserts that the constant $\gamma$ defined in~\eqref{eq:gamma}
assumes different values according to whether $G(\mathcal{H}_{\lambda},1,W(d,t))$ is in
the subcritical phrase or the supercritical phase.

\vspace{0.1in}%
\begin{lemma}\label{Lemma-Gamma}
Let $\gamma$ be defined as~\eqref{eq:gamma}. (i) If $G(\mathcal{H}_{\lambda},1,W(d,t))$
is in the subcritical phase, i.e., $\lambda<\lambda_c(\eta_1(d))$, then $\gamma<\infty$,
and $\gamma>0$ with probability 1. (ii) If $G(\mathcal{H}_{\lambda},1,W(d,t))$ is in the
supercritical phase, i.e., $\lambda>\lambda_c(\eta_1(d))$, then $\gamma=0$ with
probability 1.
\end{lemma}
\vspace{0.1in}%

\emph{Proof:} To show (i), note that $\gamma<\infty$ follows directly from
\begin{equation}\label{gamma-infinity}
\gamma=\inf_{m\geq 1}\frac{E[T_{0,m}]}{m}\leq E[T_{0,1}]<\infty,
\end{equation}
where the last inequality is shown above in the proof for
Lemma~\ref{Lemma-Limit-Convergence}.

\begin{figure}[t]
\centering
\includegraphics[width=4in]{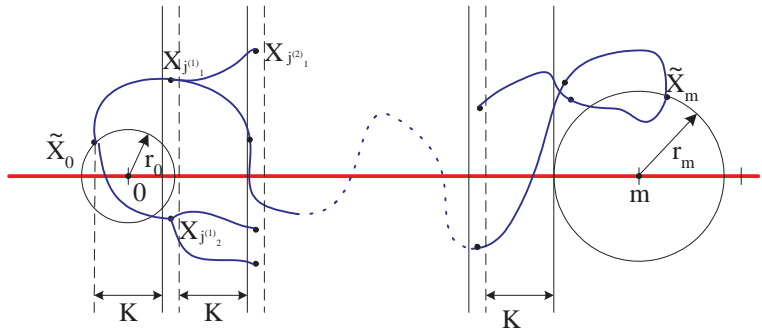}
\caption{Path segments of the paths from $\tilde{\mathbf{X}}_0$ to
$\tilde{\mathbf{X}}_m$.}\label{fig:Segment}
\end{figure}

To see why $\gamma$ is positive with probability 1, suppose the node at
$\mathbf{\tilde{X}}_0$ disseminates a message at time $t=t_0$ and consider
$G(\mathcal{H}_{\lambda},1,W(d,t_0))$. Choose $K$ large enough such that
$c_1e^{-c_2K}<\frac{1}{2}$, where $c_1$ and $c_2$ are the constants given in Proposition
\ref{Proposition-Exponential-Decay}. Let $q=\lfloor\frac{m}{2(K+1)}\rfloor$. When
$m>2(K+1)$, $q\geq 1$.

Let $S_h=\{(x,y)\in \mathbb{R}^2: K+(h-1)(K+1)\leq x-x(\mathbf{\tilde{X}}_0)<h(K+1)\}$
for $h=1,2,...$, where $x(v)$ is the $x$-coordinate of node $v$. Since
$\mathbf{\tilde{X}}_0$ and $\mathbf{\tilde{X}}_m$ are both in
$\mathcal{C}(G(\mathcal{H}_{\lambda},1))$, there exists at least one path from
$\mathbf{\tilde{X}}_0$ to $\mathbf{\tilde{X}}_m$. Moreover, since each strip $S_h$ has
width 1, at least one node of $\mathcal{C}(G(\mathcal{H}_{\lambda},1))$  lies inside each
$S_h$.

Let $\{\mathbf{X}^{(1)}_l, l=1,2,...\}$ be the nodes of
$\mathcal{C}(G(\mathcal{H}_{\lambda},1))$ which lie inside $S_1$. Since
$G(\mathcal{H}_{\lambda},1,W(d,t_0))$ is in the subcritical phase, by Proposition
\ref{Proposition-Exponential-Decay}, the probability that there exists a path consisting
of only active links from $\mathbf{\tilde{X}}_0$ to any $\mathbf{X}^{(1)}_l$,
$l=1,2,...$, is less than or equal to $c_1e^{-c_2K}<\frac{1}{2}$. In other words, with
probability strictly greater than $\frac{1}{2}$, there exists at least one inactive link
at time $t=t_0$ on any path from $\mathbf{\tilde{X}}_0$ to $\mathbf{X}^{(1)}_l$,
$l=1,2,...$. Let $T^{(1)}=\inf_{l}\{T(\mathbf{\tilde{X}}_0, \mathbf{X}^{(1)}_l)\}$. Let
$\Gamma_{W(d,t)}\triangleq \inf_{0<d\leq 1}\left\{\eta_0(d)E[Y_k(d)]\right\}>0$, then
$E[T^{(1)}]>\frac{1}{2}\Gamma_{W(d,t)}>0$.

Let $\{\mathbf{X}^{(h+1)}_{l'}, l'=1,2,...\}$ be the nodes of
$\mathcal{C}(G(\mathcal{H}_{\lambda},1))$ which lie inside $S_{h+1}$, for $h\geq 1$. By
the same argument as above, the probability that there exists a path consisting of only
active links from any node in $S_h$ to any node in $S_{h+1}$ is less than or equal to
$c_1e^{-c_2K}<\frac{1}{2}$. In other words, with probability strictly greater than
$\frac{1}{2}$, there exists at least one inactive link on any path from any node in $S_h$
to any node in $S_{h+1}$. Let $T^{(h+1)}=\inf_{l,l'}\{T(\mathbf{X}^{(h)}_l,
\mathbf{X}^{(h+1)}_{l'})\}$. Then $E[T^{(h+1)}]>\frac{1}{2}\Gamma_{W(d,t)}>0$. The path
segments are illustrated in Figure~\ref{fig:Segment}.

Since $||\mathbf{\tilde{X}}_0-\mathbf{\tilde{X}}_m||\geq m-r_0-r_m$, when
$\frac{m}{2}>r_0+r_m$, any path from $\mathbf{\tilde{X}}_0$ to $\mathbf{\tilde{X}}_m$ has
at least $\lfloor\frac{m}{2(K+1)}\rfloor=q$ segments and the delay on each segment is
strictly greater than $\frac{1}{2}\Gamma_{W(d,t)}>0$. Hence, $E[T_{0,m}]>
\frac{1}{2}q\Gamma_{W(d,t)}$ when $\frac{m}{2}>r_0+r_m$. Since both $r_0$ and $r_m$ are
finite with probability 1, $\frac{m}{2}>r_0+r_m$ holds with probability 1 as
$m\rightarrow \infty$.

Since $K$ is finite and $\Gamma_{W(d,t)}$ is positive and independent of $m$, we have
\begin{eqnarray}
\gamma
&=&\lim_{m\rightarrow\infty}\frac{E[T_{0,m}]}{m}\nonumber\\
&>& \lim_{m\rightarrow\infty}\frac{q}{m}\frac{1}{2}
\Gamma_{W(d,t)}\nonumber\\
&>&\lim_{m\rightarrow\infty}\left(\frac{1}{2(K+1)}-\frac{1}{m}\right)\frac{1}{2}
\Gamma_{W(d,t)}\nonumber\\
&>&0
\end{eqnarray}
with probability 1, where we used the fact that $q>\frac{m}{2(K+1)}-1$.

For (ii), suppose $G(\mathcal{H}_{\lambda},1,W(d,t))$ is in the supercritical phase. To
simplify notation, let $\mathcal{C}(t)$ be the infinite component of
$G(\mathcal{H}_{\lambda},1,W(d,t))$. Let $t'$ be the first time when some node in
$\mathcal{C}(t')$ receives $\mathbf{\tilde{X}}_0$'s message, and let
\[
w_1 \triangleq  \argmin_{i\in \mathcal{C}(t')} d(\mathbf{X}_i,\mathbf{\tilde{X}}_0),
\quad \mbox{and} \quad w_2 \triangleq \argmin_{i\in \mathcal{C}(t')}
d(\mathbf{X}_i,\mathbf{\tilde{X}}_m).
\]
That is, $w_1$ and $w_2$ are the nodes in the infinite component of
$G(\mathcal{H}_{\lambda},1,W(d,t'))$ with the smallest Euclidean distances to nodes
$\mathbf{\tilde{X}}_0$ and $\mathbf{\tilde{X}}_m$, respectively. If node
$\mathbf{\tilde{X}}_0$ is in $\mathcal{C}(t_0)$, then $t'=t_0$ and
$w_1=\mathbf{\tilde{X}}_0$. If at time $t'$, node $v$ is in $\mathcal{C}(t')$, then
$w_2=\mathbf{\tilde{X}}_m$.

Since both $w_1$ and $w_2$ belong to $\mathcal{C}(t')$, $T(w_1,w_2)=0$. The distances
$d(\mathbf{\tilde{X}}_0,\mathbf{X}_{w_1})$ and $d(\mathbf{X}_{w_2},\mathbf{\tilde{X}}_m)$
are finite with probability 1 by Lemma \ref{Lemma-Finite-Distance-GC} in Appendix C.
Clearly, $d(\mathbf{\tilde{X}}_0,\mathbf{X}_{w_1})$ is independent of $m$. By
stationarity, $d(\mathbf{X}_{w_2},\mathbf{\tilde{X}}_m)$ is also independent of $m$.
Hence, by the proof of Lemma~\ref{Lemma-Limit-Convergence},
$E[T(\mathbf{\tilde{X}}_0,\mathbf{X}_{w_1})]<\infty$,
$E[T(\mathbf{X}_{w_2},\mathbf{\tilde{X}}_m)]<\infty$ with probability 1 for any $m$, and
$E[T(\mathbf{\tilde{X}}_0,\mathbf{X}_{w_1})]$ and
$E[T(\mathbf{X}_{w_2},\mathbf{\tilde{X}}_m)]$ are independent of $m$. Moreover,
\begin{eqnarray}
0\leq \frac{T_{0,m}}{m} & \leq &
\frac{T(\mathbf{\tilde{X}}_0,\mathbf{X}_{w_1})+T(w_1,w_2) + T(\mathbf{X}_{w_2},\mathbf{\tilde{X}}_m)}{m}\nonumber\\
&=&\frac{T(\mathbf{\tilde{X}}_0,\mathbf{X}_{w_1})+
T(\mathbf{X}_{w_2},\mathbf{\tilde{X}}_m)}{m}.
\end{eqnarray}
Hence $\gamma=\lim_{m\rightarrow \infty}\frac{E[T_{0,m}]}{m}=0$ with probability 1. \qed

We are now ready to prove Theorem~\ref{Theorem-Delay}.

\emph{Proof of Theorem~\ref{Theorem-Delay}:} Assume node $u$ disseminates a message at
time $t=t_0$. Take $\mathbf{X}_u$ as the origin, and the line $\mathbf{X}_u\mathbf{X}_v$
as the $x$-axis. By definition $u, v\in \mathcal{C}(G(\mathcal{H}_{\lambda},1))$. Since
node $u$ is the origin, $\mathbf{X}_{u}=\tilde{\mathbf{X}}_0$. Let $m$ be the closest
integer to $x(v)$---the $x$-axis coordinate of node $\mathbf{X}_{v}$. Now
$T_{0,m}=T(\mathbf{X}_{u}, \mathbf{\tilde{X}}_m)$. If
$\mathbf{X}_{v}=\mathbf{\tilde{X}}_m$, $T(u,v)=T_{0,m}$.

Note that $m-1< d(u,v)< m+1$, Thus, for any $m>1$, we have
\begin{equation}
\frac{T_{0,m}}{m+1}< \frac{T(u,v)}{d(u,v)} < \frac{T_{0,m}}{m-1}.
\end{equation}
On the other hand, if $\mathbf{X}_{v} \neq \mathbf{\tilde{X}}_m$, then
$\mathbf{\tilde{X}}_m$ must be adjacent to $\mathbf{X}_{v}$. This is because
$||(m,0)-\mathbf{X}_{v}||\leq \frac{1}{2}$ ($m$ is the closest integer to $x(v)$) and
$||(m,0)-\mathbf{\tilde{X}}_m||\leq \frac{1}{2}$ ($\mathbf{\tilde{X}}_m$ is the closest
node to $(m,0)$). Consequently, $T_{0,m}-T(\mathbf{\tilde{X}}_m,\mathbf{X}_{v})\leq
T(u,v)\leq T_{0,m}+T(\mathbf{\tilde{X}}_m,\mathbf{X}_{v})$. Thus, for any $m>1$, we have
\begin{equation}
\frac{T_{0,m}-T(\mathbf{\tilde{X}}_m,\mathbf{X}_{v})}{m+1}< \frac{T(u,v)}{d(u,v)} <
\frac{T_{0,m}+T(\mathbf{\tilde{X}}_m,\mathbf{X}_{v})}{m-1}.
\end{equation}

Since $\mathbf{\tilde{X}}_m$ is adjacent to $\mathbf{X}_{v}$,
$T(\mathbf{\tilde{X}}_m,\mathbf{X}_{v})< \infty$ with probability 1. Therefore, in both
cases, by Lemma~\ref{Lemma-Limit-Convergence} and a typical $\epsilon$-$\delta$ argument
(see Appendix D), we have for any $\epsilon>0, \delta>0$, there exists $d_0<\infty$, such
that if $d(u,v)>d_0$, then
\begin{equation}
\Pr\left(\left|\frac{T(u,v)}{d(u,v)} -\gamma\right|<\epsilon\right)>1-\delta.
\end{equation}
When $G(\mathcal{H}_{\lambda},1)$ is in the subcritical phase, by
Lemma~\ref{Lemma-Gamma}, we have $0<\gamma<\infty$ with probability 1.

On the other hand, when $G(\mathcal{H}_{\lambda},1)$ is in the supercritical phase, by
Lemma~\ref{Lemma-Gamma}, we have $\gamma=0$ with probability 1. Then, by a typical
$\epsilon$-$\delta$ argument (see Appendix E), we have for any $\epsilon>0, \delta>0$,
there exists $d_0<\infty$, such that if $d(u,v)>d_0$ then
\[
\Pr\left(\frac{T(u,v)}{d(u,v)}<\epsilon\right)>1-\delta.
\]
\qed

\subsection{Effects of Propagation Delay}

Up to this point, we have ignored propagation delays. We now take this type of delay into
account.  Suppose the propagation delay is $0<\tau<\infty$ for any link, independent of
the link length. We assume the following mechanism is used for a transmission from node
$i$ to node $j$: (i) a packet is successfully received by node $j$ if the length of the
active period on link $(i,j)$, during which the packet is being transmitted, is greater
than or equal to $\tau$; (ii) node $i$ retransmits a packet to node $j$ until the packet
is successfully received by $j$.

Note that due to the Markovian nature of the link state processes $\{W_{ij}(d_{ij},t)\}$,
at the instant when a packet arrives at node $i$, the residual active time for link
$(i,j)$ has the same distribution as $Z(d_{ij})$. Thus without loss of generality, we assume
that node $i$ initiates transmission on link $(i,j)$ at time $0$. If link $(i,j)$ is on at
time 0 with $Z_1(d)\geq \tau$, then the transmission delay $T^{\tau}_{ij}(d)$ on $(i,j)$
is $\tau$. However, if link $(i,j)$ is on at time 0 with $Z_1(d)< \tau$, or if $(i,j)$ is
off at time $t =0$, then the delay on $(i,j)$ is less straightforward to calculate. In
this case, we need to capture the behavior of retransmissions. Let
\begin{equation}\label{eq:K}
K(d)=\argmin_{k\geq 1}\{Z_k(d)\geq\tau\}.
\end{equation}
Then, $K(d)$ is a stopping time for the sequence $\{Z_k(d), k \geq 1\}$.  Now we have
\begin{equation}\label{eq:T-ij}\left\{
\begin{array}{lll}
T^{\tau}_{ij} =\displaystyle \sum_{i=1}^{K-1}(Y_i+Z_i)+Y_K+\tau, & W(d,0)=0,\\
T^{\tau}_{ij} =\displaystyle \sum_{i=1}^{K-1}(Y_i+Z_i)+\tau, &W(d,0)=1,
\end{array}\right.
\end{equation}
where we abbreviate $T^{\tau}_{ij}(d)$, $K(d)$, $Y_i(d)$ and $Z_i(d)$ as $T^{\tau}_{ij}$,
$K$, $Y_i$ and $Z_i$, respectively.

Let
\begin{equation}
T^{\tau}(u,v)=T^{\tau}(\mathbf{X}_u,\mathbf{X}_v)\triangleq\inf_{l(u,v)\in
\mathcal{L}(u,v)}\left\{\sum_{(i,j) \in l(u,v)}T^{\tau}_{ij}(d_{ij})\right\},
\end{equation}
where $T^{\tau}_{ij}(d_{ij})$ is given by~\eqref{eq:T-ij}. Then, $T^{\tau}(u,v)$ is the
message delay on the path from $u$ to $v$ with the smallest delay, including propagation
delays.

\vspace{+.1in}%
\begin{corollary}\label{Corollary-Propagation-Delay}
Given $G(\mathcal{H}_{\lambda},1,W(d,t))$ with $\lambda>\lambda_c$ and propagation delay
$0<\tau<\infty$, there exists a constant $\gamma(\tau)<\infty$ with
$\gamma(\tau)\geq\tau$ (with probability 1), such that for any $u,v \in
\mathcal{C}(G(\mathcal{H}_{\lambda},1))$, and any $\epsilon>0,\delta>0$, there exists
$d_0<\infty$ such that for any $u,v$ with $d(u,v)>d_0$,
\begin{equation}\label{LinearRelation-propagation-delay}
\Pr\left(\left|\frac{T^{\tau}(u,v)}{d(u,v)}-\gamma(\tau)\right|<\epsilon\right)>1-\delta.
\end{equation}
Moreover, when $G(\mathcal{H}_{\lambda},1,W(d,t))$ is in the subcritical phase, as
$\tau\rightarrow 0$, $\gamma(\tau)\rightarrow\gamma$ with probability 1, where $\gamma$
is defined in Theorem \ref{Theorem-Delay}. When $G(\mathcal{H}_{\lambda},1,W(d,t))$ is in
the supercritical phase, as $\tau\rightarrow 0$, $\gamma(\tau)\rightarrow 0$ with
probability 1.
\end{corollary}
\vspace{+.1in}%

To prove this corollary, we need the following two lemmas.

\vspace{0.1in}%
\begin{lemma}\label{Lemma-Propogation-delay}
Given any $0<\tau<\infty$, for all $0<d\leq 1$, the expected delay on each link $(i,j)$
is positive and finite, i.e.,
\begin{equation}
0<E[T^{\tau}_{ij}]<\infty.
\end{equation}
\end{lemma}
\vspace{0.1in}%

\emph{Proof:} By~\eqref{eq:T-ij}, we have
\begin{eqnarray}\label{eq:E-T-ij}
E[T^{\tau}_{ij}]&=&E[E[T^{\tau}_{ij}|W(d,0)]]\nonumber\\
&=&\eta_0E[T^{\tau}_{ij}|W(d,0)=0]+\eta_1E[T^{\tau}_{ij}|W(d,0)=1]\nonumber\\
&=&\eta_0E\left[\sum_{i=1}^{K-1}(Y_i+Z_i)+Y_K+\tau|Z_i<\tau, i=1,...,K-1\right] \nonumber \\
& &  +\eta_1E\left[\sum_{i=1}^{K-1}(Y_i+Z_i)+\tau|Z_i<\tau, i=1,...,K-1\right]\nonumber\\
&=&\tau+\eta_0E[Y_K]+E\left[\sum_{i=1}^{K-1}(Y_i+Z_i)|Z_i<\tau,i=1,...,K-1\right]\nonumber\\
&<& E[K]\tau+\eta_0E[Y_K]+(E[K]-1)E[Y_i],
\end{eqnarray}
where in the last equality, we used the fact that $Y_i$ and $Z_i$ are i.i.d. and
$Z_i<\tau$ for $i=1,2,...K-1$, as well as Wald's Equality for stopping time $K$.

Since $0<\tau<\infty$, $0<\eta_0<1$, and $0<E[Y_i]<\infty$, in order to show
$0<E[T^{\tau}_{ij}]<\infty$, it suffices to show $1\leq E[K]<\infty$. By definition,
$K\geq 1$ so that $E[K]\geq 1$. Thus, we need only to show $E[K]<\infty$. For any $k\geq
1$, $\Pr(K=k)=\Pr(Z_1<\tau,...,Z_{k-1}<\tau,Z_k\geq \tau)=F_Z(\tau)^{k-1}(1-F_Z(\tau))$,
where $F_Z(\cdot)=\Pr(Z_i \leq \tau)$. Then
\begin{equation}\label{eq:E-K}
E[K]=\sum_{k=1}^{\infty}kF_Z(\tau)^{k-1}(1-F_Z(\tau))=\frac{1}{1-F_Z(\tau)}.
\end{equation}
Therefore, we have $E[K]<\infty$. \qed

\vspace{0.1in}%
\begin{lemma}\label{Lemma-Shotest-Path}
Given $G(\mathcal{H}_{\lambda},1,W(d,t))$ with $\lambda>\lambda_c$ and no propagation
delay, let $L_{0,m}$ be the path from $\mathbf{\tilde{X}_0}$ to $\mathbf{\tilde{X}_m}$
that attains $T_{0,m}$ and has the smallest number of links (in case there exist multiple
paths attaining $T_{0,m}$). Then $|L_{0,m}|<\infty$ with probability 1 for each $m$,
where $|L_{0,m}|$ is the number of links along $L_{0,m}$.
\end{lemma}
\vspace{0.1in}%

\emph{Proof:} By the proof of Lemma~\ref{Lemma-Finite-Hops}, we have $E[T_{0,m}]<\infty$.
We can express $E[T_{0,m}]$ as
\[
E[T_{0,m}]=E[E[T_{0,m}||L_{0,m}|]],
\]
where
\[
E[T_{0,m}||L_{0,m}|]=\sum_{i=1}^{|L_{0,m}|}\eta_0^{(i)}(d)E[Y_k^{(i)}(d)]\geq
|L_{0,m}|\Gamma_{W(d,t)},
\]
where $\eta_0^{(i)}(d)$ and $E[Y_k^{(i)}(d)]$ are the stationary probability of the
inactive state, and the expected inactive period of the $i$-th link with length $d$ on
$L_{0,m}$ respectively, and $\Gamma_{W(d,t)}=\inf_{0<d\leq 1}\{\eta_0(d)E[Y_k(d)]\}>0$.
Thus, we have
\[
E[|L_{0,m}|]\Gamma_{W(d,t)}<\infty.
\]
This implies $E[|L_{0,m}|]<\infty$, which further implies $|L_{0,m}|<\infty$ with
probability 1. \qed

\emph{Proof of Corollary~\ref{Corollary-Propagation-Delay}:} Let
$T^{\tau}_{l,m}=T^{\tau}(\mathbf{\tilde{X}}_l,\mathbf{\tilde{X}}_m)$, for
$||\mathbf{\tilde{X}}_l-\mathbf{\tilde{X}}_m||<\infty$, $0\leq l\leq m$, where
$\mathbf{\tilde{X}}_i$ is defined as in~\eqref{X-i-tilde}.

Clearly, the relationship $T^{\tau}_{0,m}\leq T^{\tau}_{0,l}+T^{\tau}_{l,m}$ still holds
for any $0\leq l\leq m$. Hence, condition (i) of
Theorem~\ref{Theorem-Subadditive-Ergodic} holds. Since the propagation delay does not
affect the stationarity of the geometric structure of the network, conditions (ii) and
(iii) of Theorem~\ref{Theorem-Subadditive-Ergodic} also hold.

By the same argument as that in the proof of Lemma~\ref{Lemma-Finite-Hops}, we have
$E[|L|]<\infty$, where $|L|\triangleq|L(\mathbf{\tilde{X}}_0,\mathbf{\tilde{X}}_m)|$ and
$L(\mathbf{\tilde{X}}_0,\mathbf{\tilde{X}}_m)$ is the shortest path from
$\mathbf{\tilde{X}}_0$ to $\mathbf{\tilde{X}}_m$. Let $T^{\tau,L}_{0,m}$ be the delay on
this path. Then,
\[
E[T^{\tau,L}_{0,m}||L|]=\sum_{i=1}^{|L|}E[T^{\tau}_i(d_i)]\leq
|L|\Lambda_{W^{\tau}(d,t)},
\]
where $T^{\tau}_i(d_i)$ is the delay on the $i$-th link with length $d_i$ on the path
$L(\mathbf{\tilde{X}}_0,\mathbf{\tilde{X}}_m)$, as given by~\eqref{eq:T-ij}, and
$\Lambda_{W^{\tau}(d,t)}\triangleq \sup_{0<d\leq 1}E[T^{\tau}_i(d_i)]<\infty$. By
Lemma~\ref{Lemma-Propogation-delay}, we have $0<E[T^{\tau}_i(d_i)]<\infty$ for all
$0<d_i\leq 1$, so that $\Lambda_{W^{\tau}(d,t)}<\infty$. Hence
\[
E[T^{\tau,L}_{0,m}]=E[E[T^{\tau,L}_{0,m}||L|]]\leq E[|L|]\Lambda_{W^{\tau}(d,t)}<\infty,
\]
which implies $E[T^{\tau}_{0,m}]<\infty$. This ensures that condition (iv) of
Theorem~\ref{Theorem-Subadditive-Ergodic} holds.

Furthermore, the propagation delay does not affect the strong mixing property of
$\{T^{\tau}_{l,m},0\leq l\leq m\}$. Therefore the result of
Lemma~\ref{Lemma-Limit-Convergence} holds for $\{T^{\tau}_{l,m},0\leq l\leq m\}$. Let
$\gamma(\tau)\triangleq \lim_{m\rightarrow\infty}\frac{E[T^{\tau}_{0,m}]}{m}$, then
$\gamma(\tau)=\inf_{m\geq 1}\frac{E[T^{\tau}_{0,m}]}{m}$, and
\begin{equation}\label{eq:converge}
\lim_{m\rightarrow\infty}\frac{T^{\tau}_{0,m}}{m}=\gamma(\tau) \quad \mbox{with
probability 1}.
\end{equation}

Then applying the same proof for Theorem~\ref{Theorem-Delay}, we can show that for any
$\epsilon>0, \delta>0$, there exists $d_0<\infty$, such that if $d(u,v)>d_0$, then
\[
\Pr\left(\left|\frac{T^{\tau}(u,v)}{d(u,v)}
-\gamma(\tau)\right|<\epsilon\right)>1-\delta.
\]

To see why $\gamma(\tau)<\infty$ and $\gamma(\tau)\geq \tau$ with probability 1, first
note that
\begin{equation}
\gamma(\tau)=\inf_{m\geq 1}\frac{E[T^{\tau}_{0,m}]}{m}\leq E[T^{\tau}_{0,1}]<\infty.
\end{equation}
Moreover, since the shortest path between nodes $\mathbf{\tilde{X}}_0$ and
$\mathbf{\tilde{X}}_m$ has at least
$\lfloor||\mathbf{\tilde{X}}_0-\mathbf{\tilde{X}}_m||\rfloor\geq \lfloor
m-r_0-r_m\rfloor$ links, $T^{\tau}_{0,m}\geq\tau \lfloor m-r_0-r_m\rfloor$. Since $r_0$
and $r_m$ are both finite with probability 1 and independent of $m$, we have
$\gamma(\tau)\geq \tau$ with probability 1.

In the following, we show that as $\tau\rightarrow 0$, $\gamma(\tau)\rightarrow\gamma$
with probability 1 when $G(\mathcal{H}_{\lambda},1)$ is in the subcritical phase, and
$\gamma(\tau)\rightarrow 0$ with probability 1 when $G(\mathcal{H}_{\lambda},1)$ is in
the supercritical phase. Observe that
\[
T_{0,m}\leq T^{\tau}_{0,m}\leq \sum_{i=1}^{|L_{0,m}|}T^{\tau}_i(d_i),
\]
where $L_{0,m}$ is defined in Lemma~\ref{Lemma-Shotest-Path}, and $T^{\tau}_i(d_i)$ is
the delay on the $i$-th link with length $d_i$ along $L_{0,m}$, as given
by~\eqref{eq:T-ij}. From Lemma~\ref{Lemma-Shotest-Path}, we have $|L_{0,m}|<\infty$ with
probability 1. Thus with probability 1,
\[
E[T_{0,m}]\leq E[T^{\tau}_{0,m}]\leq \sum_{i=1}^{|L_{0,m}|}E[T^{\tau}_i(d_i)].
\]
By~\eqref{eq:E-T-ij} and $E[T_{0,m}]=\sum_{i=1}^{|L_{0,m}|}\eta_0(d_i)E[Y_k(d_i)]$ we
have
\begin{equation}\label{eq:T-p-0-m}
E[T_{0,m}]\leq E[T^{\tau}_{0,m}]\leq E[T_{0,m}]+|L_{0,m}|E[K]\tau +
\sum_{i=1}^{|L_{0,m}|}(E[K]-1)E[Y_k(d_i)],
\end{equation}
with probability 1. From~\eqref{eq:E-K}, we know that as $\tau\rightarrow 0$,
$E[K]\rightarrow 1$. Therefore, as $\tau\rightarrow 0$, we have $|L_{0,m}|E[K]\tau +
\sum_{i=1}^{|L_{0,m}|}(E[K]-1)E[Y_k(d_i)]\rightarrow 0$ with probability 1. This,
combined with~\eqref{eq:T-p-0-m} implies $\lim_{\tau\rightarrow 0} E[T^{\tau}_{0,m}]=
E[T_{0,m}]$ with probability 1. Therefore,
\begin{eqnarray}
\lim_{\tau\rightarrow 0}\gamma(\tau)&=& \lim_{\tau\rightarrow 0}\lim_{m\rightarrow
\infty}\frac{E[T^{\tau}_{0,m}]}{m}\nonumber\\
&=& \lim_{m\rightarrow \infty} \lim_{\tau\rightarrow 0} \frac{E[T^{\tau}_{0,m}]}{m}\nonumber\\
&=& \lim_{m\rightarrow \infty} \frac{E[T_{0,m}]}{m}\nonumber\\
&=& \gamma,
\end{eqnarray}
with probability 1, where the interchanging of limitation operations is justified by
$E[T^{\tau}_{0,m}]<\infty$. Consequently, as $\tau\rightarrow 0$,
$\gamma(\tau)\rightarrow\gamma$ with probability 1 when $G(\mathcal{H}_{\lambda},1)$ is
in the subcritical phase. Since $\gamma \rightarrow 0$ with probability 1 if
$G(\mathcal{H}_{\lambda},1)$ is in the supercritical phase, we have
$\gamma(\tau)\rightarrow 0$ with probability 1 in this case. \qed

An interesting observation of this corollary is when the propagation delay is large, the
message delay cannot be improved too much by transforming the network from the
subcritical phase to the supercritical phase. However, as the propagation delay becomes
negligible, the message delay scales almost sub-linearly ($\gamma(\tau)\approx 0$) when
the network is in the supercritical phase, while the delay scales linearly
($\gamma(\tau)\approx \gamma$) when the network is in the subcritical phase.

\section{Numerical Experiments}

\begin{figure}[t!]
\centerline{ \subfigure[Subcritical]{
\includegraphics[width=2.5in]{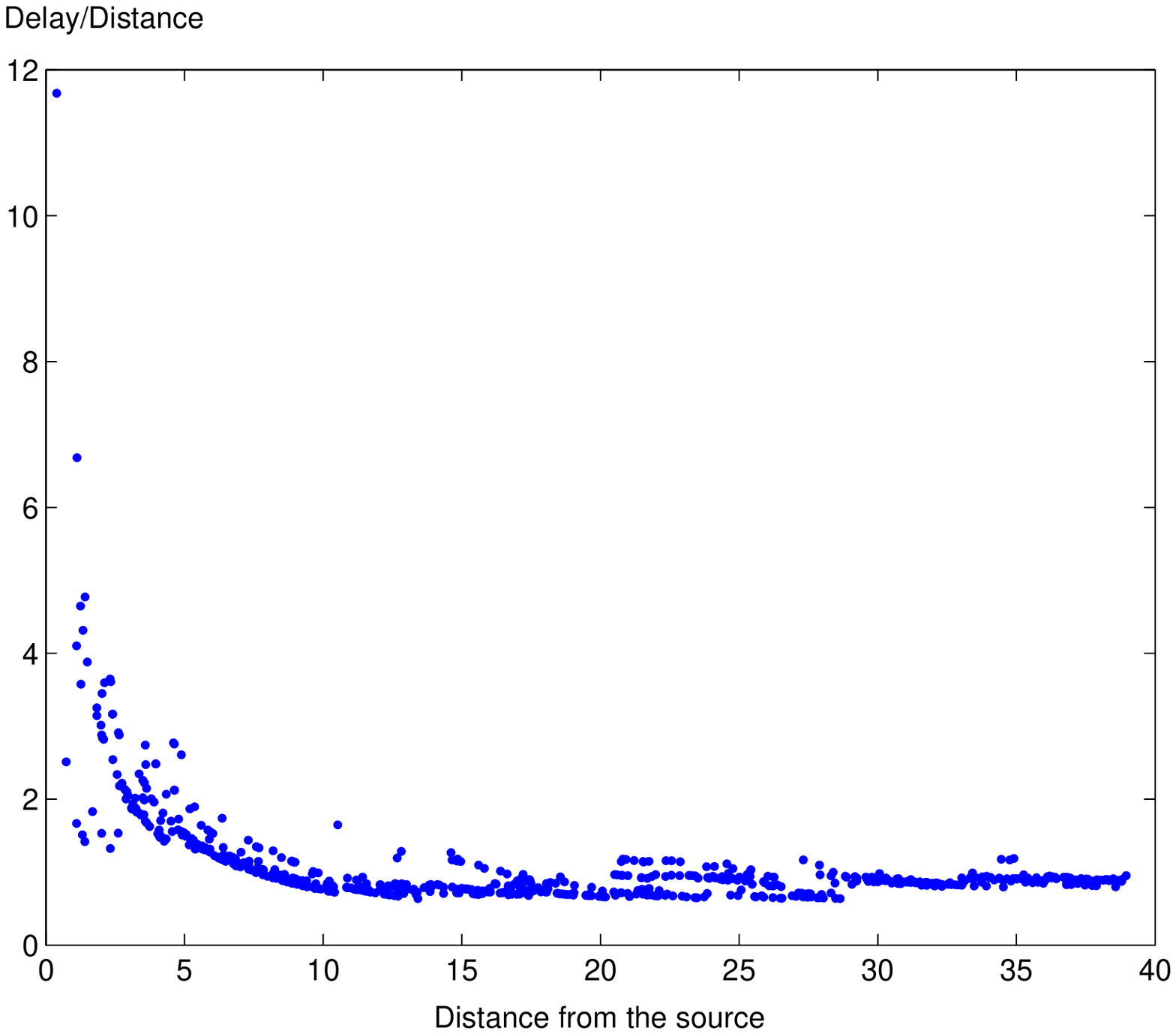}}\hfil
\subfigure[Supercritical]{
\includegraphics[width=2.5in]{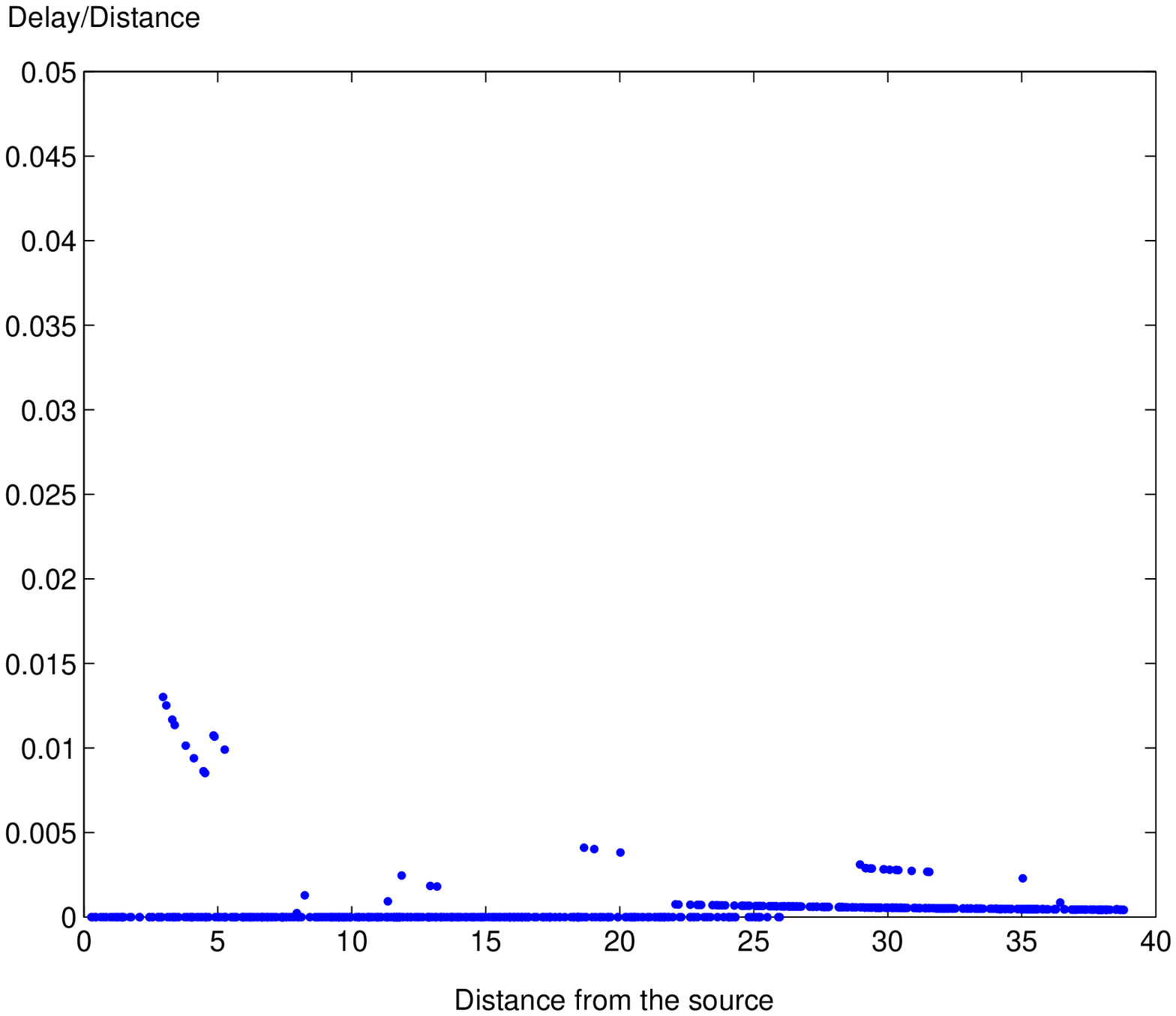}}}
\caption{Delay performance of information dissemination in wireless networks with dynamic
unreliable links ($\lambda=1.75$): (a) $E[T_1(d)]=0.5$ and $E[T_0(d)]=2$ for any $0<d\leq
1$; (b) $E[T_1(d)]=2.5$ and $E[T_0(d)]=0.5$ for any $0<d\leq
1$.}\label{fig:Latency-Exp-Inde}
\end{figure}

\begin{figure}[t!]
\centerline{ \subfigure[Subcritical]{
\includegraphics[width=2.5in]{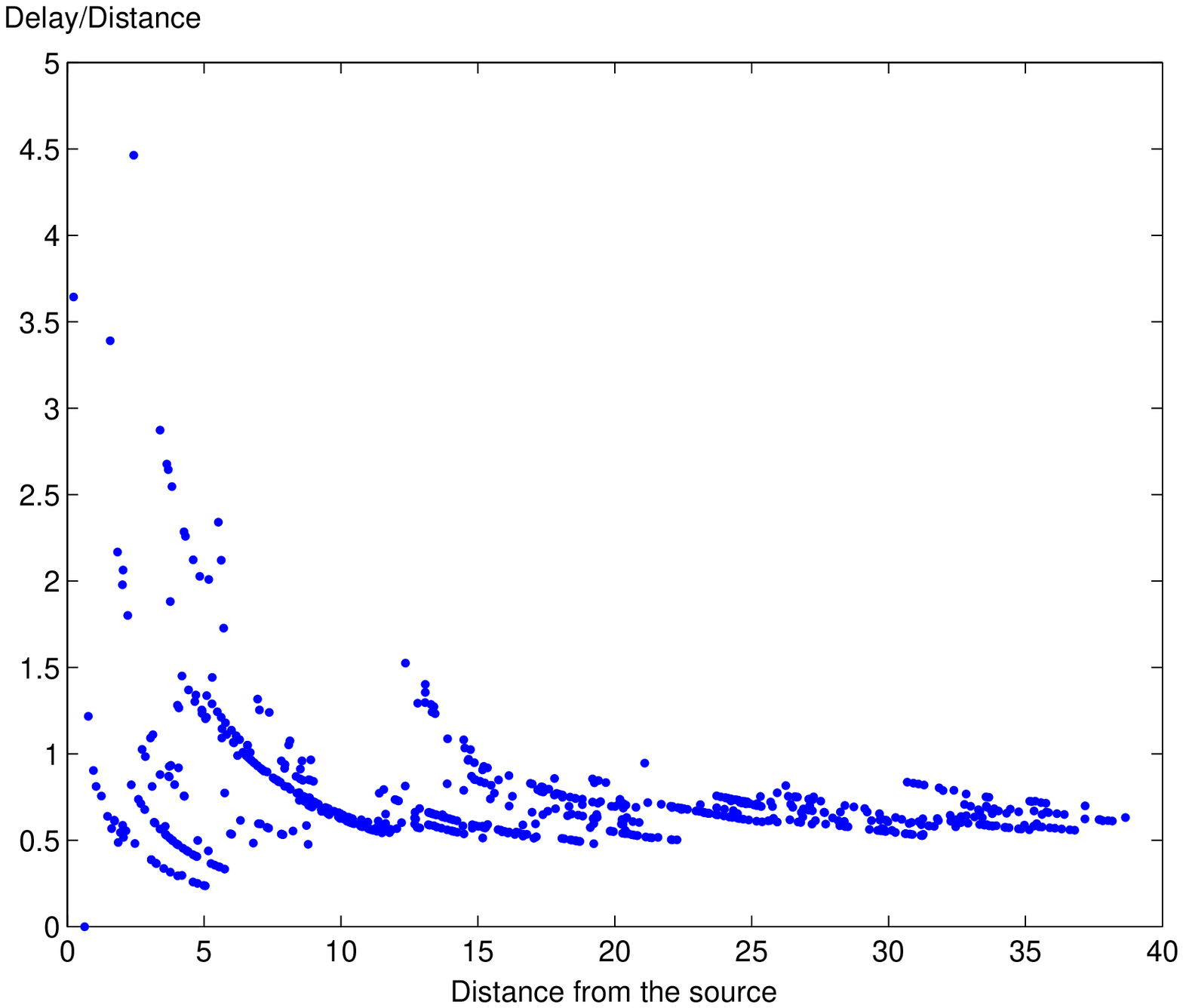}}\hfil
\subfigure[Supercritical]{
\includegraphics[width=2.5in]{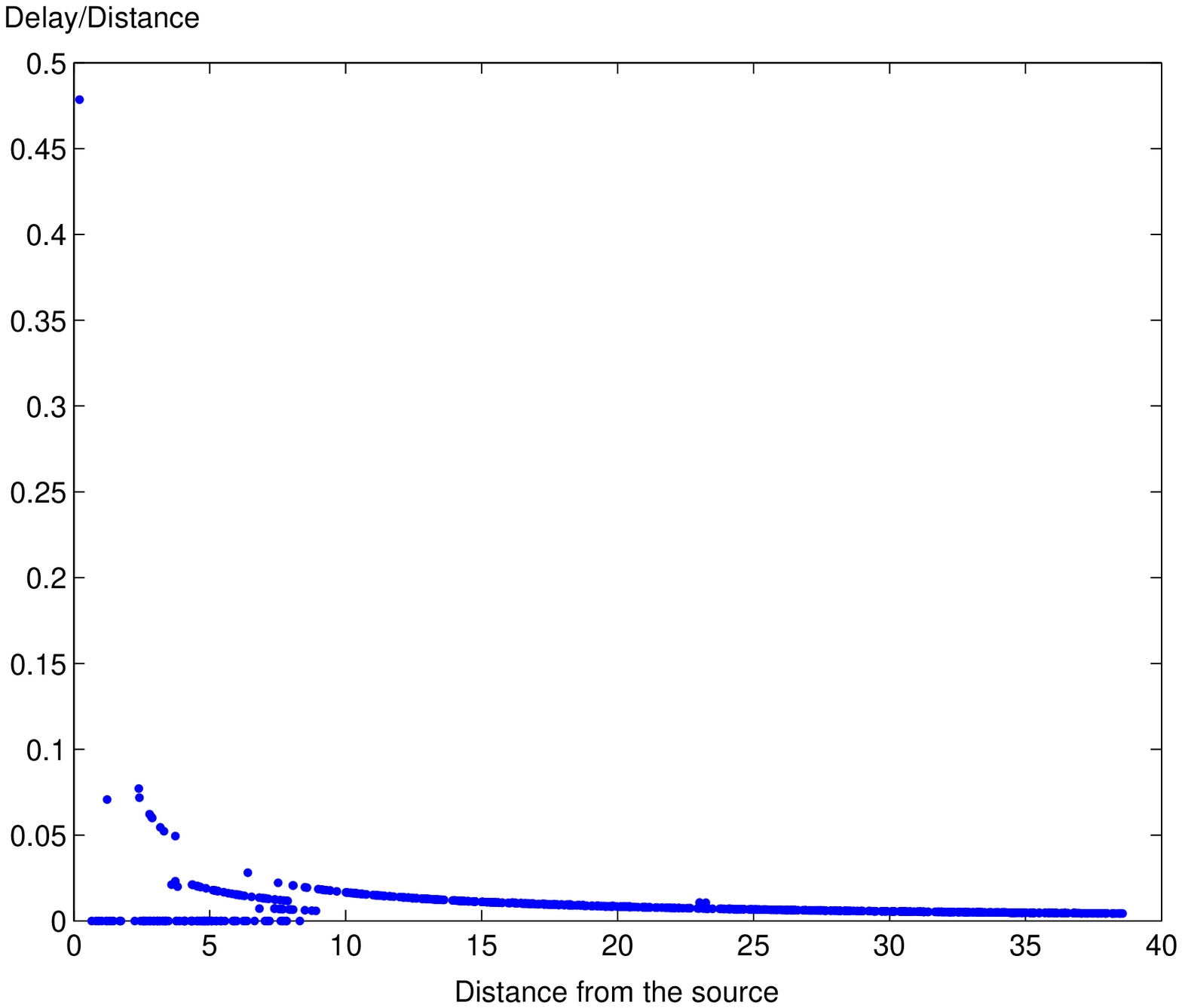}}}
\caption{Delay performance of information dissemination in wireless networks with dynamic
unreliable links ($\lambda=1.875$): (a) $E[T_1(d)]=0.5$ and $E[T_0(d)]=1.5d+1$ for any
$0<d\leq 1$; (b) $E[T_1(d)]=2$ and $E[T_0(d)]=0.5d+0.5$ for any $0<d\leq
1$.}\label{fig:Latency-Exp-De}
\end{figure}

\begin{figure}[t!]
\centerline{ \subfigure[Subcritical]{
\includegraphics[width=2.5in]{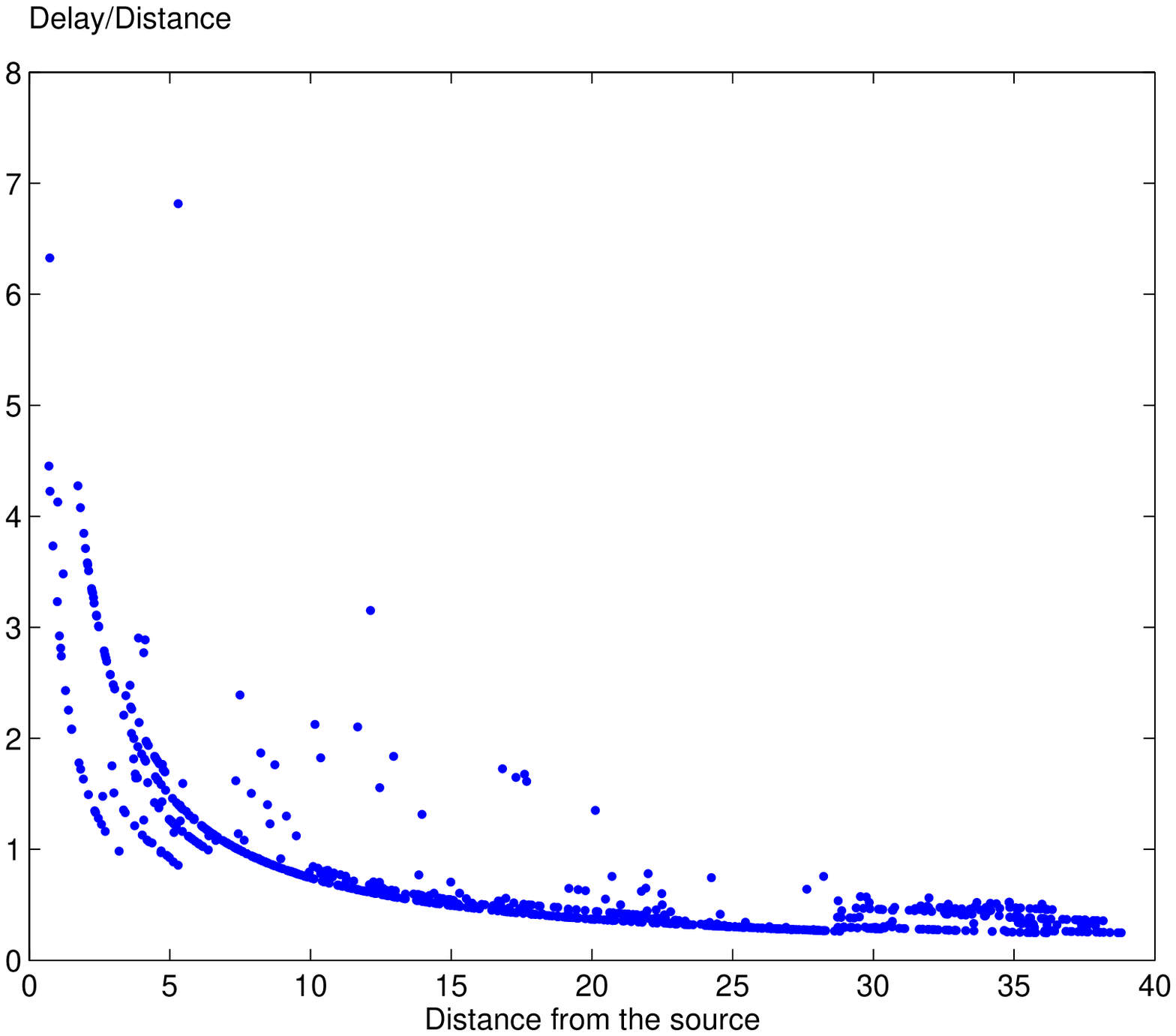}}\hfil
\subfigure[Supercritical]{
\includegraphics[width=2.5in]{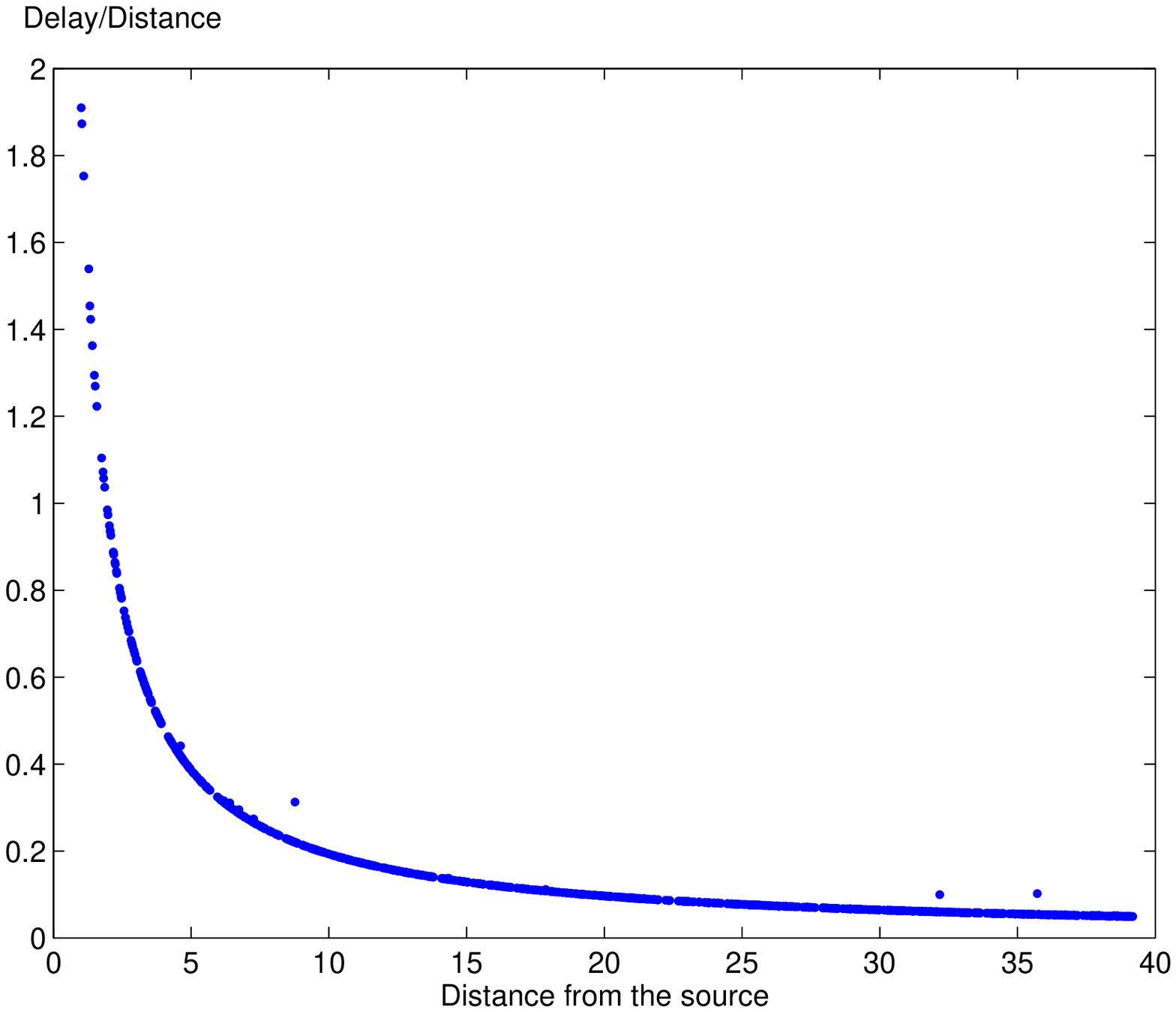}}}
\caption{Delay performance of information dissemination in wireless networks with dynamic
unreliable links ($\lambda=1.875$) and propagation delay $\tau=1$: (a) $E[T_1(d)]=1$ and
$E[T_0(d)]=8$ for any $0<d\leq 1$; (b) $E[T_1(d)]=1$ and $E[T_0(d)]=2$ for any $0<d\leq
1$.}\label{fig:Latency-Exp-Prga}
\end{figure}

In this section, we present some simulation results.
Figure~\ref{fig:Latency-Exp-Inde}-\ref{fig:Latency-Exp-Prga} show simulation results of
the information dissemination delay performance in large-scale wireless networks with
dynamic unreliable links.

In Figure~\ref{fig:Latency-Exp-Inde}, the lengths of the active and inactive periods have
exponential distributions independent of $d$---the length of the link. In
Figure~\ref{fig:Latency-Exp-De}, the lengths of the active and inactive periods have
exponential distributions depending on $d$. In all of these scenarios, it can be seen
that when the resulting dynamic network is in the subcritical phase,
$\frac{T(u,v)}{d(u,v)}$ converges to a non-zero value as $d(u,v)\rightarrow\infty$.  The
limit depends on the density of $G(\mathcal{H}_{\lambda},1)$ and the distributions and
expected values of the active and inactive periods. When the resulting dynamic network is
in the supercritical phase, $\frac{T(u,v)}{d(u,v)}$ converges to zero as
$d(u,v)\rightarrow\infty$.

To see how propagation delays affect the message delay, and to verify the results of
Corollary \ref{Corollary-Propagation-Delay}, we illustrate simulation results in
Figure~\ref{fig:Latency-Exp-Prga}, where $T_1(d)$ and $T_0(d)$ have exponential
distributions independent of $d$.

\section{Conclusions}

In this paper, we studied percolation-based connectivity and information dissemination
latency in large-scale wireless networks with unreliable links. We first studied static
models, where each link of the network is functional (or active) with some probability,
independently of all other links. We then studied wireless networks with dynamic
unreliable links, where each link is active or inactive according to Markov on-off
processes. We showed that a phase transition exists in such dynamic networks, and the
critical density for this model is the same as the corresponding one for static networks
(under some mild conditions). We further investigated the delay performance in such
networks by modelling the problem as a first passage percolation process on random
geometric graphs. We showed that without propagation delay, the delay of information
dissemination scales linearly with the Euclidean distance between the sender and the
receiver when the resulting network is in the subcritical phase, and the delay scales
sub-linearly with the distance if the resulting network is in the supercritical phase. We
further showed that when propagation delay is taken into account, the delay of
information dissemination always scales linearly with the Euclidean distance between the
sender and the receiver.

\section*{Appendix A}

\emph{Proof of Proposition~\ref{Proposition-Exponential-Decay}:} Let $B$ be a bounded box
containing the origin, and let $W(B)$ be the union of components that have some node(s)
of $G(\mathcal{H}_{\lambda}, 1, p_e(\cdot))$ inside box $B$. Precisely,
$W(B)=\{\mbox{component } W'\in G(\mathcal{H}_{\lambda}, 1, p_e(\cdot)): \exists w\in W',
\mathbf{x}_w\in B\}$.

Consider the following two events:
\[
E\triangleq\{d(W(B))\geq h\}, \quad \mbox{and}\quad F\triangleq\{\mbox{all nodes of
$G(\mathcal{H}_{\lambda}, 1, p_e(\cdot))$ inside $B$ belong to $W_{\mathbf{0}}$}\}.
\]
Clearly, events $E$ and $F$ are both increasing events. By the FKG inequality, we have
$\Pr(E\cap F)\geq\Pr(E)\Pr(F)$. Thus,
\begin{eqnarray}\label{eq:diameter-relationship}
\Pr(d(W_{\mathbf{0}})\geq h) & \geq & \Pr(E\cap F)\nonumber\\
&\geq & \Pr(E)\Pr(F)\nonumber\\
&= & \Pr(F)\Pr(d(W(B))\geq h),
\end{eqnarray}
where $\Pr(F)>0$ since $B$ is bounded. By~\eqref{eq:diameter-relationship}, we have
\[
E[d(W(B)]\leq \frac{E[d(W_{\mathbf{0}})]}{\Pr(F)}.
\]
Therefore, when $\lambda<\lambda_c(p_e(\cdot))$, we have $E[d(W_{\mathbf{0}})]<\infty$
and thus $E[d(W(B)]<\infty$.

To prove the Proposition, it is sufficient to show $\Pr(B \leftrightsquigarrow
B(h)^c)\leq c_1 e^{-c_2h}$, where $\{B \leftrightsquigarrow B(h)^c\}$ denotes the event
that some node(s) inside $B$ and some nodes in $B(h)^c$ are connected.

We partition the space as the union of
$B(i,j)\triangleq\left(i-\frac{1}{2},i+\frac{1}{2}\right]\times
\left(j-\frac{1}{2},j+\frac{1}{2}\right]$, where $(i,j)\in \mathbb{Z}^2$.  Since
$E[d(W(B(0,0))]<\infty$, $d(W(B(0,0))<\infty$ with probability 1. Then we can choose $M$
sufficiently large so that $E[H_M]<\frac{1}{6}$, where $H_M$ is the number of boxes
$B(i,j)$ outside $B(M)=[-M,M]^2$ intersecting $W(B(0,0))$.

\begin{figure}[t!]
\centering
\includegraphics[width=3in]{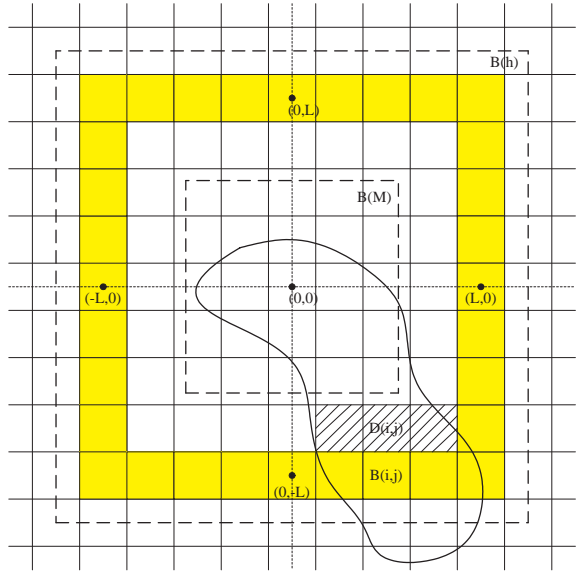}
\caption{$B(h)$, $B(M)$, $L$, $B(i,j)$ and $D(i,j)$.}\label{fig:Box}
\end{figure}

Now choose $L$ large enough so that the set $\bigcup_{m(i,j)\geq L-1}B(i,j)$ is disjoint
from $B(M)$, where $m(i,j)=\max\{|i|,|j|\}$. Choose $h$ sufficient large so that
$\bigcup_{m(i,j)\leq L}B(i,j)\subset B(h)$. Observe that if $\{B(0,0)\leftrightsquigarrow
B(h)^c\}$ occurs, then there exists $(i,j)$ with $m(i,j)=L$ for which
$\{B(0,0)\leftrightsquigarrow D(i,j)\}$  and $\{B(i,j)\leftrightsquigarrow B(h)^c\}$
occur disjointly,\footnote{Let $U$ be a bounded Borel set in $\mathbb{R}^2$. For any
realization $G\in G(H_{\lambda},1,p_e(\cdot))$, let $G_u=(V_u,E_u)$, where $V_u=\{v: v\in
G\cap U\}$ and $E_u=\{(u,v):u,v\in V_u\}$. Define $[G_u]=\{G'\in
G(H_{\lambda},1,p_e(\cdot)): \exists G''\subset G' \mbox{ s.t. } G''_u=G_u\}$. We say
that an increasing event $A$ is an event on $U$ if $I_A(G)=1$ and $G'\in[G_u]$ imply that
$I_A(G')=1$. A rational rectangle is an open 2-dimensional box with rational coordinates.
Let $A$ and $B$ be two increasing events on $U$, and $W_1$ and $W_2$ be two disjoint sets
that are finite unions of rational rectangles. For $G\in G(H_{\lambda},1,p_e(\cdot))$, if
$I_A(G'_{W_1})=1$ where $G'_{W_1}\in[G_{W_1}]$, and $I_B(G'_{W_2})=1$ where
$G'_{W_2}\in[G_{W_2}]$, then we say that $A$ and $B$ occur disjointly. We use $A\Box B$
to denote the event that $A$ and $B$ occur disjointly. For details, please refer
to~\cite{MeRo96, Gr99}.}, where $D(i,j)\triangleq \bigcup_{m(i'j,')=L-1,
m(i-i',j-j')=1}B(i',j')$. This is illustrated in Figure~\ref{fig:Box}.

Let $\{B(0,0) \leftrightsquigarrow D(i,j)\Box B(i,j) \leftrightsquigarrow B(h)^c\}$
denote the event that $\{B(0,0) \leftrightsquigarrow D(i,j)\}$ and $\{B(i,j)
\leftrightsquigarrow B(h)^c\}$ occur disjointly. It then follows from the BK
inequality~\cite{MeRo96, Gr99} that
\begin{eqnarray}
\Pr(B(0,0) \leftrightsquigarrow B(h)^c) & \leq & \sum_{(i,j):m(i,j)=L}\Pr(B(0,0)
\leftrightsquigarrow D(i,j)\Box B(i,j) \leftrightsquigarrow B(h)^c)\nonumber\\
&\leq &\max_{(i,j):m(i,j)=L}\Pr(B(i,j) \leftrightsquigarrow
B(h)^c)\sum_{(i,j):m(i,j)=L}\Pr(B(0,0)
\leftrightsquigarrow D(i,j))\nonumber\\
&=& \max_{(i,j):m(i,j)=L}\Pr(B(i,j) \leftrightsquigarrow B(h)^c)
\sum_{(i,j):m(i,j)=L}E[I_{\{B(0,0)
\leftrightsquigarrow D(i,j)\}}] \nonumber\\
&=& \max_{(i,j):m(i,j)=L}\Pr(B(i,j) \leftrightsquigarrow
B(h)^c)E\left[\sum_{(i,j):m(i,j)=L}I_{\{B(0,0)
\leftrightsquigarrow D(i,j)\}}\right] \nonumber\\
&\leq& \max_{(i,j):m(i,j)=L}\Pr(B(i,j) \leftrightsquigarrow B(h)^c)3E[H_M],
\end{eqnarray}
where the last inequality follows from the fact the each box $B(i',j')$ can be contained
in at most 3 $D(i,j)$'s.

It follows that
\begin{equation}\label{eq:Pr-B-0-0}
\Pr(B(0,0) \leftrightsquigarrow B(h)^c) \leq \frac{1}{2} \max_{(i,j):m(i,j)=L}\Pr(B(i,j)
\leftrightsquigarrow B(h)^c).
\end{equation}
To bound the right hand side of~\eqref{eq:Pr-B-0-0}, choose a sufficiently large $h$ such
that $\bigcup_{m(i'-i,j'-j)=L, m(i,j)=L}B(i',j')\subset B(h)$. The same argument as above
shows that for all $(i,j)$ with $m(i,j)=L$,
\begin{equation}
\Pr(B(i,j) \leftrightsquigarrow B(h)^c) \leq \frac{1}{2}
\max_{(i',j'):m(i'-i,j'-j)=L}\Pr(B(i',j') \leftrightsquigarrow B(h)^c).
\end{equation}
Repeating this argument leads to the desired conclusion. \qed

\section*{Appendix B}

The following lemma is similar to the one used in \cite{DoFrTh05,DoFrMaMeTh06,Gr99}. For
completeness, we provide the proof here.

\vspace{0.1in}%
\begin{lemma}\label{Lemma-Closed-Circuit-Number}
Given a square lattice $\mathcal{L}'$, suppose that the origin is located at the center
of one square. Let the number of circuits\footnote{A circuit in a lattice $\mathcal{L}'$
is a closed path with no repeated vertices in $\mathcal{L}'$.} surrounding the origin
with length $2m$ be $\gamma(2m)$, where $m\geq 2$ is an integer, then we have
\begin{equation}\label{gamma-2m}
\gamma(2m)\leq \frac{4}{27}(m-1)3^{2m}.
\end{equation}
\end{lemma}
\vspace{0.1in}%
\begin{figure}[t!]
\centering
\includegraphics[width=2.5in]{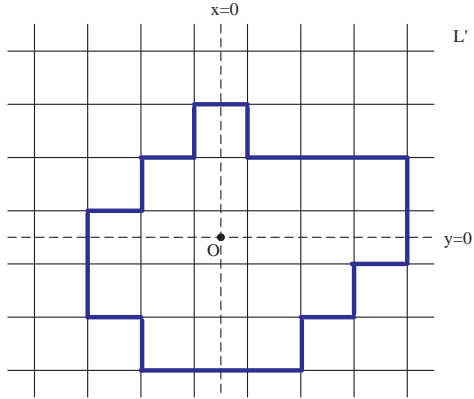}
\caption{An example of a circuit surrounding the origin in lattice
$\mathcal{L}'$}\label{fig:ClosedCircuit}
\end{figure}

\emph{Proof:} In Figure~\ref{fig:ClosedCircuit}, an example of a circuit that surrounds
the origin is illustrated. First note that the length of such a circuit must be even.
This is because there is a one-to-one correspondence between each pair of edges above and
below the line $y=0$, and similarly for each pair of edges at the left and right of the
line $x=0$. Furthermore, the rightmost edge can be chosen only from the lines $l_i:
x=i-\frac{1}{2}, i=1,...,m-1$. Hence the number of possibilities for this edge is at most
$m-1$. Because this edge is the rightmost edge, each of the two edges adjacent to it has
two choices for its direction. For all the other edges, each one has at most three
choices for its direction. Therefore the number of total choices for all the other edges
is at most $3^{2m-3}$. Consequently, the number of circuits that surround the origin and
have length $2m$ must be less or equal to $(m-1)2^23^{2m-3}$, and hence we have
(\ref{gamma-2m}). \qed

\section*{Appendix C}

\begin{lemma}\label{Lemma-Finite-Distance-GC}
Suppose $G(\mathcal{H}_{\lambda},1,p_e(\cdot))$ is in the supercritical phase, i.e,
$\lambda>\lambda_c(p_e(\cdot))$. Let $v\notin
\mathcal{C}(G(\mathcal{H}_{\lambda},1,p_e(\cdot)))$ and define
\[
w\triangleq \argmin_{i\in \mathcal{C}(G(\mathcal{H}_{\lambda},1,p_e(\cdot)))} d(i,v),
\]
i.e., $w$ is the node in the infinite component of
$G(\mathcal{H}_{\lambda},1,p_e(\cdot))$ with the smallest Euclidean distances to node
$v$. Then, $d(w,v)<\infty$ with probability 1.
\end{lemma}
\vspace{+0.1in}%

The idea behind the proof for this lemma is similar to that for the proof for
Lemma~\ref{Lemma-Finite-Distance}. The difference is that the probability of a good event
is now defined with respect to $G(\mathcal{H}_{\lambda},1,p_e(\cdot))$ instead of
$G(\mathcal{H}_{\lambda},1)$.

Given $G(\mathcal{H}_{\lambda},1,p_e(\cdot))$ with $\lambda>\lambda_c(p_e(\cdot))$, as in
the proof for Lemma~\ref{Lemma-Finite-Distance}, we consider a mapping between
$G(\mathcal{H}_{\lambda},1,p_e(\cdot))$ and a square lattice
$\mathcal{L}=d\cdot\mathbb{Z}^2$, where $d$ is the edge length. The vertices of
$\mathcal{L}$ are located at $(d\times i, d\times j)$ where $(i,j)\in \mathbb{Z}^2$. For
each horizontal edge $a$, let the two end vertices be $(d\times a_x, d\times a_y)$ and
$(d\times a_x+d, d\times a_y)$.

As in the proof for Lemma~\ref{Lemma-Finite-Distance}, define event $A_a(d,p_e(\cdot))$
for edge $a$ in $\mathcal{L}$ as the set of outcomes for which the following condition
holds: The rectangle $R_a=[a_xd-\frac{d}{4},a_xd+\frac{5d}{4}]\times
[a_yd-\frac{d}{4},a_yd+\frac{d}{4}]$ is crossed from \emph{left to right} by a connected
component in $G(\mathcal{H}_\lambda,1,p_e(\cdot))$. Define event $A_a'(d,p_e(\cdot))$ for
edge $a$ in $\mathcal{L}$ as the set of outcomes for which the following condition holds:
The rectangle $R_a=[a_xd-\frac{d}{4},a_xd+\frac{5d}{4}]\times
[a_yd-\frac{d}{4},a_yd+\frac{d}{4}]$ is crossed from \emph{top to bottom} by a connected
component in $G(\mathcal{H}_\lambda,1,p_e(\cdot))$.

Let
\begin{equation}\label{eq:p-g-d-pe}
p_g(d,p_e(\cdot))\triangleq\Pr(A_a(d,p_e(\cdot))), \quad \mbox{and} \quad
p_g'(d,p_e(\cdot))\triangleq\Pr(A_a'(d,p_e(\cdot))).
\end{equation}
Define $A_a(d,p_e(\cdot))$ and $A_a'(d,p_e(\cdot))$ similarly for all vertical edges by rotating
the rectangle by $90^{\circ}$.

Define a \emph{vacant component} $V$ in $\mathbb{R}^2$ with respect to (w.r.t.)
$G(\mathcal{H}_{\lambda},1,p_e(\cdot))$ to be a region $V\subset \mathbb{R}^2$ such that
$V\cap G(\mathcal{H}_{\lambda},1,p_e(\cdot))=\emptyset$ (i.e., no node or any part of a
link of $G(\mathcal{H}_{\lambda},1,p_e(\cdot))$ is contained in $V$), and such that there
exists no other region $U\subset \mathbb{R}^2$ satisfying $V\subset U$ and $U\cap
G(\mathcal{H}_{\lambda},1,p_e(\cdot))=\emptyset$.

\vspace{0.1in}%
\begin{definition} For $G(\mathcal{H}_{\lambda},1,p_e(\cdot))$, let $V_{\mathbf{0}}$ be
the vacant component in $\mathbb{R}^2$ w.r.t. $G(\mathcal{H}_{\lambda},1,p_e(\cdot))$
containing $\mathbf{0}$. Let
\begin{equation}
\lambda_c^*(p_e(\cdot))\triangleq\sup \{\lambda: \Pr(d(V_{\mathbf{0}})=\infty)>0\}.
\end{equation}
\end{definition}
\vspace{0.1in}%

Similarly we can define the vacant component $V_{\mathbf{0}}'$ containing the origin in
$\mathbb{R}^2$ w.r.t. $G(\mathcal{H}_{\lambda},1)$, and $\lambda_c^*\triangleq\sup
\{\lambda: \Pr(d(V_{\mathbf{0}}')=\infty)>0\}$. It is known that $\lambda_c^*=\lambda_c$
(Chapter 4 in~\cite{MeRo96}). Since $G(\mathcal{H}_{\lambda},1, p_e(\cdot))$ is a
subgraph of $G(\mathcal{H}_{\lambda},1)$, it is clear that $\lambda_c^*(p_e(\cdot))\geq
\lambda_c^*$.

\vspace{0.1in}%
\begin{proposition}\label{Proposition-Zero-One}
Let $\psi^*(p_e(\cdot))\triangleq \Pr(\exists \mbox{ vacant component }V\subset
\mathbb{R}^2\mbox{ w.r.t. } G(\mathcal{H}_{\lambda},1,p_e(\cdot)): d(V)=\infty)$. Then
\begin{equation}\label{eq:psi-pe}
\psi^*(p_e(\cdot))=\left\{\begin{array}{ll}1, & \lambda<\lambda_c^*(p_e(\cdot)),\\0, &
\lambda>\lambda_c^*(p_e(\cdot)).
\end{array}\right.
\end{equation}
\end{proposition}
\vspace{0.1in}%

\emph{Proof:} First assume $\lambda<\lambda_c^*(p_e(\cdot))$. The graph
$G(\mathcal{H}_{\lambda},1, p_e(\cdot))$ is obtained by placing a link between two nodes
$i$ and $j$ with probability $p_e(\cdot)$ when $||\mathbf{x}_i-\mathbf{x}_j||\leq 1$. The
event $\{\exists \mbox{ vacant component }V\subset \mathbb{R}^2\mbox{ w.r.t. }
G(\mathcal{H}_{\lambda},1,p_e(\cdot)): d(V)=\infty\}$ does not depend on the existence of
any finite collection of those links. By Kolmogorov's zero-one law~\cite{Gr99, Du96},
$\psi^*(p_e(\cdot))$ assumes the values 0 and 1 only. Since
$\Pr(d(V_{\mathbf{0}})=\infty)>0$, then
\[
\psi^*(p_e(\cdot))\geq \Pr(d(V_{\mathbf{0}})=\infty)>0,
\]
so that $\psi^*(p_e(\cdot))=1$ by Kolmogorov's zero-one law.

On the other hand, if $\lambda>\lambda_c^*(p_e(\cdot))\geq\lambda_c$, with probability 1,
there is no vacant component with infinite diameter in $\mathbb{R}^2$ w.r.t.
$G(\mathcal{H}_{\lambda},1)$ (Chapter 4 in~\cite{MeRo96}).  Since
$\Pr(d(V_{\mathbf{0}})=\infty)=0$, we have
\[
\psi^*(p_e(\cdot))\leq \sum_{\mathbf{x}\in \mathbb{Q}^2}\Pr(d(V_{\mathbf{x}})=\infty)=0,
\]
where we used the fact that $\mathbb{Q}^2$ is dense and any infinite vacant component is
open so that any infinite component contains at least one $\mathbf{x}\in
\mathbb{Q}^2$.\qed

Given the mapping between $G(\mathcal{H}_{\lambda},1,p_e(\cdot))$ and $\mathcal{L}$,
define event $A_a^*(d,p_e(\cdot))$ for edge $a$ in $\mathcal{L}$ as the set of outcomes
for which the following condition holds: the rectangle
$R_a=[a_xd-\frac{d}{4},a_xd+\frac{5d}{4}]\times [a_yd-\frac{d}{4},a_yd+\frac{d}{4}]$ is
crossed from \emph{left} to \emph{right} by a \emph{vacant} component in $\mathbb{R}^2$
w.r.t. $G(\mathcal{H}_\lambda,1,p_e(\cdot))$. Define event $A_a^{*'}(d,p_e(\cdot))$ for
edge $a$ in $\mathcal{L}$ as the set of outcomes for which the following condition holds:
the rectangle $R_a=[a_xd-\frac{d}{4},a_xd+\frac{5d}{4}]\times
[a_yd-\frac{d}{4},a_yd+\frac{d}{4}]$ is crossed from \emph{top} to \emph{bottom} by a
\emph{vacant} component in $\mathbb{R}^2$ w.r.t. $G(\mathcal{H}_\lambda,1,p_e(\cdot))$.

\begin{figure}[t]
\centering
\includegraphics[width=4in]{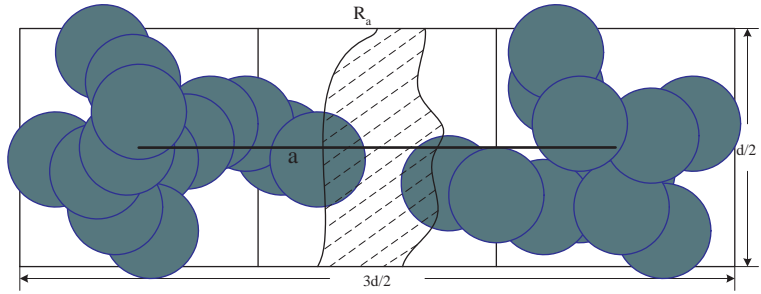}
\caption{The rectangle $R_a$ is crossed from \emph{top} to \emph{bottom} by a
\emph{vacant} component in $\mathbb{R}^2$ w.r.t.
$G(\mathcal{H}_\lambda,1,p_e(\cdot))$}\label{fig:VacantRectangle}
\end{figure}

Let
\begin{equation}\label{eq:p-g-d-t}
p_g^*(d,p_e(\cdot))\triangleq\Pr(A_a^*(d,p_e(\cdot))),\quad\mbox{and}\quad
p_g^{*'}(d,p_e(\cdot))\triangleq\Pr(A_a^{*'}(d,p_e(\cdot))).
\end{equation}
Define $A_a^*(d,p_e(\cdot))$ and $A_a^{*'}(d,p_e(\cdot))$ similarly for all vertical edges by
rotating the rectangle by $90^{\circ}$. Figure~\ref{fig:VacantRectangle} illustrates
$A_a^{*'}(d,p_e(\cdot))$.

We now define another critical density with respect to
$G(\mathcal{H}_\lambda,1,p_e(\cdot))$.

\vspace{0.1in}%
\begin{definition} Given $G(\mathcal{H}_{\lambda},1,p_e(\cdot))$, let
\begin{equation}
\lambda_S^*(p_e(\cdot))\triangleq\sup \{\lambda:
\lim\sup_{d\rightarrow\infty}p_g^{*'}(d,p_e(\cdot))>0\}.
\end{equation}
\end{definition}

\vspace{0.1in}%
\begin{proposition}\label{Proposition-Vacant-Critical-Identity}
For $G(\mathcal{H}_{\lambda},1,p_e(\cdot))$, we have
\begin{equation}\label{eq:vacant-critical-idnetity}
\lambda_c(p_e(\cdot))=\lambda_c^*(p_e(\cdot))=\lambda_S^*(p_e(\cdot)).
\end{equation}
\end{proposition}
\vspace{0.1in}%

\emph{Proof:} To show~\eqref{eq:vacant-critical-idnetity}, it is sufficient to show (i)
$\lambda_c(p_e(\cdot))\leq \lambda_c^*(p_e(\cdot))$, (ii) $\lambda_c^*(p_e(\cdot))\leq
\lambda_S^*(p_e(\cdot))$, and (iii) $\lambda_S^*(p_e(\cdot))\leq \lambda_c(p_e(\cdot))$.

To show (i) $\lambda_c(p_e(\cdot))\leq \lambda_c^*(p_e(\cdot))$, let
$\lambda<\lambda_c(p_e(\cdot))$. Then $G(\mathcal{H}_{\lambda},1,p_e(\cdot))$ is in the
subcritical phase. Let $B_1(i)=(0,2i)+B(1)$ where $B(1)=[-1,1]^2$ for $i=0,1,2...$.
Observe that the existence of a left to right crossing in rectangle
$[0,3^k]\times[0,3^{k+1}]$ by a component $W'$ of $G(\mathcal{H}_{\lambda},1,p_e(\cdot))$
implies the existence of a component $W''$ of $G(\mathcal{H}_{\lambda},1,p_e(\cdot))$
starting from $\bigcup_{i=0}^{\lceil\frac{3^{k+1}}{2}\rceil}B_1(i)$ (i.e., the first node
in $W''$ in the x-axis direction is inside
$\bigcup_{i=0}^{\lceil\frac{3^{k+1}}{2}\rceil}B_1(i)$) with diameter greater than or
equal to $3^k-2$. Hence, we have for any $k\geq 1$,
\begin{eqnarray}
p_g'(d=2\cdot3^k,p_e(\cdot))&\leq& \Pr\left(\bigcup_{i=0}^{\lceil\frac{3^{k+1}}{2}\rceil}
\{d(W(B_1(i)))\geq 3^k-2\}\right)\nonumber\\
&\leq & \bigcup_{i=0}^{\lceil\frac{3^{k+1}}{2}\rceil}\Pr(d(W(B_1(i)))\geq 3^k-2)\nonumber\\
&=& \left(\left\lceil\frac{3^{k+1}}{2}\right\rceil+1\right)\Pr(d(W(B(1)))\geq 3^k-2)\nonumber\\
&<& \left(\frac{9}{2}3^{k-1}+2\right)\Pr(d(W(B(1)))\geq 3^k-2)\nonumber\\
&\leq& \left(\frac{9}{2}3^{k-1}+2\right)\Pr(d(W(B(1)))\geq 3^{k-1}),
\end{eqnarray}
where $W(B_1(i))$ is the union of components of $G(\mathcal{H}_{\lambda}, 1, p_e(\cdot))$
that have some node(s) inside box $B_1(i)$. Precisely, $W(B_1(i))=\{\mbox{component } W'
\mbox{ of } G(\mathcal{H}_{\lambda}, 1, p_e(\cdot)): \exists w\in W', \mathbf{x}_w\in
B_1(i)\}$.

Since $\lambda<\lambda_c(p_e(\cdot))=\lambda_D(p_e(\cdot))$, $E[d(W_0)]<\infty$. By the same
argument used in the proof for Proposition~\ref{Proposition-Exponential-Decay}, we have
$E[d(W(B(1))]<\infty$.

Let $P_k=\Pr(d(W(B(1)))\geq k)$. Then $P_k$ is non-increasing in $k$, and thus we have
\begin{eqnarray}
\sum_{k=1}^{\infty}p_g'(d=2\cdot3^k,p_e(\cdot))&<&
\sum_{k=1}^{\infty}\left(\frac{9}{2}3^{k-1}+2\right)P_{3^{k-1}}\nonumber\\
&=& \sum_{k=0}^{\infty}\left(\frac{9}{2}3^k+2\right)P_{3^k}\nonumber\\
&=& \frac{9}{2}\sum_{k=0}^{\infty}3^kP_{3^k}+2
\sum_{k=0}^{\infty}P_{3^k}\nonumber\\
&\leq& \frac{9}{2}\left(P_1+3\sum_{k=1}^{\infty}3^{k-1}P_{3^k}\right)+2E[d(W(B(1)))]\nonumber\\
&\leq& \frac{9}{2}(P_1+3E[d(W(B(1)))])+2E[d(W(B(1)))]\nonumber\\
&<&\infty.
\end{eqnarray}

Note that $p_g'(d=2\cdot3^k,p_e(\cdot))+p_g^*(d=2\cdot3^k,p_e(\cdot))=1$ for all $k\geq 1$. Hence
by the Borel-Cantelli Lemma, we have
\[
\Pr(\exists \mbox{ vacant top to bottom crossing $t_k$ in } [0,3^k]\times [0,3^{k+1}] \mbox{ for
all suffcient large }k)=1.
\]
Rotational invariance implies that
\[
\Pr(\exists \mbox{ vacant left to right crossing $l_k$ in } [0,3^{k+2}]\times [0,3^{k+1}] \mbox{
for all suffcient large }k)=1.
\]

\begin{figure}[t]
\centering
\includegraphics[width=4in]{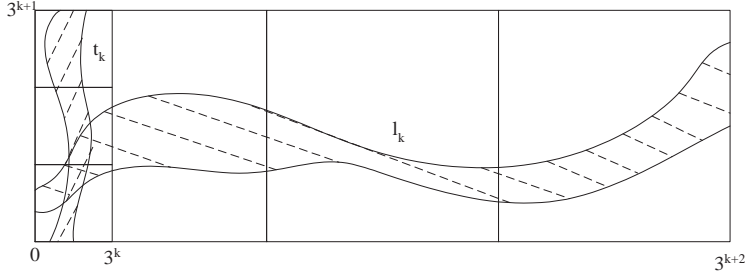}
\caption{A vertical crossing $t_k$ of $[0,3^k]\times [0,3^{k+1}]$ and a horizontal
crossing $l_k$ of $[0,3^{k+2}]\times [0,3^{k+1}]$ must
intersect.}\label{fig:VacantComponent}
\end{figure}

As illustrated in Figure~\ref{fig:VacantComponent}, a vertical crossing $t_k$ of
$[0,3^k]\times [0,3^{k+1}]$ and a horizontal crossing $l_k$ of $[0,3^{k+2}]\times
[0,3^{k+1}]$ must intersect. Also, $t_{k+1}$ of $[0,3^{k+1}]\times [0,3^{k+2}]$ and $l_k$
must intersect. Thus the union of vacant crossings $\{t_k\}$ and $\{l_k\}$ combines to
give an infinite vacant component in the first quadrant. Therefore, by
Proposition~\ref{Proposition-Zero-One}, $\lambda\leq \lambda_c^*(p_e(\cdot))$, and
$\lambda_c(p_e(\cdot))\leq \lambda_c^*(p_e(\cdot))$.

We now show (ii) $\lambda_c^*(p_e(\cdot))\leq \lambda_S^*(p_e(\cdot))$. Let
$\lambda>\lambda_S^*(p_e(\cdot))$. Then $\lim\sup_{d\rightarrow\infty}p_g^{*'}(d,p_e(\cdot))=0$,
and hence $\lim\sup_{d\rightarrow\infty}p_g(d,p_e(\cdot))=1$. Then there exists $\delta>0$ such
that there are infinitely many $d_1',d_2',...$ satisfying $p_g(d_i',p_e(\cdot))\geq\delta$ for
$i=1,2,...$. Now choose $d_1=d_1'$ and $d_{i+1}=\min\{d_j':d_j'\geq 3d_i'\}$. Then by the same
argument used in the proof for Lemma~\ref{Lemma-Finite-Hops}, we can construct infinitely many
annuli around the origin, each annulus having edge length $d_i'$ and containing a circuit with a
probability larger than $\delta$. Then, by the Borel-Cantelli Lemma, with probability 1, there
exist infinitely many circuits surrounding the origin and hence $d(V_{\mathbf{0}})$ is finite with
probability 1. This implies that $\lambda>\lambda_c^*(p_e(\cdot))$, and thus
$\lambda_S^*(p_e(\cdot))\geq \lambda_c^*(p_e(\cdot))$.

Finally, (iii) $\lambda_S^*(p_e(\cdot))\leq \lambda_c(p_e(\cdot))$ can be shown by the same
argument as that for the proof of Theorem 4.3 and Theorem 4.4 in~\cite{MeRo96}. \qed

\emph{Proof of Lemma~\ref{Lemma-Finite-Distance-GC}:} If $G(\mathcal{H}_{\lambda},1,p_e(\cdot))$
is in the supercritical phase,
$\lambda>\lambda_c(p_e(\cdot))=\lambda_c^*(p_e(\cdot))=\lambda_S^*(p_e(\cdot))$. Thus,
$\lim\sup_{d\rightarrow\infty}p_g^{*'}(d,p_e(\cdot))=0$ and
$\lim\sup_{d\rightarrow\infty}p_g(d,p_e(\cdot))=1$. Then by the same methods used in the proof for
Lemma~\ref{Lemma-Finite-Distance}, we can show Lemma~\ref{Lemma-Finite-Distance-GC}. \qed

\section*{Appendix D}

Since $T(\mathbf{\tilde{X}}_m,\mathbf{X}_v)<\infty$ with probability 1, for any
$0<\delta_1<\delta$, there exists $M<\infty$ such that
\[
\Pr(T(\mathbf{\tilde{X}}_m,\mathbf{X}_v)<M)>1-\delta_1.
\]

Then for any $\epsilon>0$,
\begin{eqnarray*}\label{eq:epsilon-delta-1}
\Pr\left(\left|\frac{T(u,v)}{d(u,v)}-\gamma\right|<\epsilon\right) &= &
\Pr\left(\gamma-\epsilon<\frac{T(u,v)}{d(u,v)}<\gamma+\epsilon\right)\nonumber\\
&\geq&\Pr\left(\gamma-\epsilon<\frac{T(u,v)}{d(u,v)}<\gamma+\epsilon|
T(\mathbf{\tilde{X}}_m,\mathbf{X}_v)<M\right)\Pr(T(\mathbf{\tilde{X}}_m,\mathbf{X}_v)<M)\nonumber\\
&>&\Pr\left(\gamma-\epsilon<\frac{T(u,v)}{d(u,v)}<\gamma+\epsilon|
T(\mathbf{\tilde{X}}_m,\mathbf{X}_v)<M\right)(1-\delta_1)\nonumber\\
&\geq&\Pr\left(\gamma-\epsilon<\frac{T_{0,m}-M}{m+1}, \frac{T_{0,m}+M}{m-1}<\gamma+\epsilon\right)
(1-\delta_1)\nonumber\\
&=& \Pr\big((\gamma-\epsilon+M)+(\gamma-\epsilon)m<T_{0,m}<
m(\gamma+\epsilon)-(M+\gamma+\epsilon)\big)(1-\delta_1)\nonumber\\
&\geq& \Pr\big((\gamma+\epsilon+M)+(\gamma-\epsilon)m<T_{0,m}<
m(\gamma+\epsilon)-(M+\gamma+\epsilon)\big)(1-\delta_1).
\end{eqnarray*}

Since $\lim_{m\rightarrow\infty}\frac{T_{0,m}}{m}=\gamma$ with probability 1, for
$\delta_2=1-\frac{1-\delta}{1-\delta_1}$, there exists $m_0<\infty$ such that for any
$m>m_0$,
\[
\Pr\left(\gamma-\frac{\epsilon}{2}<\frac{T_{0,m}}{m}<\gamma+\frac{\epsilon}{2}\right)>1-\delta_2.
\]
If $\gamma-\frac{\epsilon}{2}<\frac{T_{0,m}}{m}<\gamma+\frac{\epsilon}{2}$, then
\[
T_{0,m}<\left(\gamma+\frac{\epsilon}{2}\right)m<m(\gamma+\epsilon)-(M+\gamma+\epsilon),
\]
and
\[
T_{0,m}>\left(\gamma-\frac{\epsilon}{2}\right)m>m(\gamma-\epsilon)+(M+\gamma+\epsilon).
\]
Hence, for any $m>\max\{m_0,\frac{2(M+\gamma+\epsilon)}{\epsilon}\}$, we have
\[
\Pr\left((\gamma+\epsilon+M)+(\gamma-\epsilon)m<T_{0,m}<
m(\gamma+\epsilon)-(M+\gamma+\epsilon)\right)>1-\delta_2.
\]
Moreover, since $m>d(u,v)-1$, if $d(u,v)>d_0\triangleq
\max\{m_0,\frac{2(M+\gamma+\epsilon)}{\epsilon}\}+1$, we have
$m>\max\{m_0,\frac{2(M+\gamma+\epsilon)}{\epsilon}\}$, so that
\[
\Pr\left(\left|\frac{T(u,v)}{d(u,v)}-\gamma\right|<\epsilon\right)>(1-\delta_1)(1-\delta_2)=1-\delta.
\]

\section*{Appendix E}

Let $\epsilon>0$, $0<\delta<1$ be given. When $G(\mathcal{H}_{\lambda},1,W(d,t))$ is in
the supercritical phase, $\gamma=0$ with probability 1. Thus, there exists
$0<\epsilon_1<\epsilon$ and $0<\delta_1<\delta$ such that
\[
\Pr(\gamma<\epsilon_1)>1-\delta_1.
\]

Let $\epsilon_2=\epsilon-\epsilon_1$, and $\delta_2=1-\frac{1-\delta}{1-\delta_1}$. From
Appendix D, we know that for $\epsilon_2$ and $\delta_2$, there exist $d_0<\infty$ such
that when $d(u,v)>d_0$,
\[
\Pr\left(\gamma-\epsilon_2<\frac{T(u,v)}{d(u,v)}<\gamma+\epsilon_2\right)>1-\delta_2.
\]

Then for the given $\epsilon$, when $d(u,v)>d_0$, we have
\begin{eqnarray*}
\Pr\left(\frac{T(u,v)}{d(u,v)}<\epsilon\right) &\geq &
\Pr\left(\frac{T(u,v)}{d(u,v)}<\epsilon|\gamma+\epsilon_2<\epsilon\right)
\Pr(\gamma+\epsilon_2<\epsilon)\\
&>&\Pr\left(\frac{T(u,v)}{d(u,v)}<\epsilon|\gamma+\epsilon_2<\epsilon\right)(1-\delta_1)\\
&\geq& \Pr\left(\frac{T(u,v)}{d(u,v)}<\gamma+\epsilon_2\right)(1-\delta_1)\\
&>&(1-\delta_2)(1-\delta_1)\\
&=&1-\delta.
\end{eqnarray*}

\bibliography{PercolationTopic2,DisseminationMobility}

\begin{thebibliography}{10}

\bibitem{GuKu98}
P.~Gupta and P.~R. Kumar, ``Critical power for asymptotic connectivity in
  wireless networks,'' in {\em Stochastic Analysis, Control, Optimization and
  Applications: A Volume in Honor of W. H. Fleming}, pp.~547--566, 1998.

\bibitem{Gi61}
E.~N. Gilbert, ``Random plane networks,'' {\em J. Soc. Indust. Appl. Math.},
  vol.~9, pp.~533--543, 1961.

\bibitem{Gr99}
G.~Grimmett, {\em Percolation}.
\newblock New York: Springer, second~ed., 1999.

\bibitem{MeRo96}
R.~Meester and R.~Roy, {\em Continuum Percolation}.
\newblock New York: Cambridge University Press, 1996.

\bibitem{Pe03}
M.~Penrose, {\em Random Geometric Graphs}.
\newblock New York: Oxford University Press, 2003.

\bibitem{BoNrFrMe03}
L.~Booth, J.~Bruck, M.~Franceschetti, and R.~Meester, ``Covering algorithms,
  continuum percolation and the geometry of wireless networks,'' {\em Annals of
  Applied Probability}, vol.~13, pp.~722--741, May 2003.

\bibitem{FrBoCoBrMe05}
M.~Franceschetti, L.~Booth, M.~Cook, J.~Bruck, and R.~Meester, ``Continuum
  percolation with unreliable and spread out connections,'' {\em Journal of
  Statistical Physics}, vol.~118, pp.~721--734, Feb. 2005.

\bibitem{DoMaTh04}
O.~Dousse, P.~Mannersalo, and P.~Thiran, ``Latency of wireless sensor networks
  with uncoordinated power saving mechniasm,'' in {\em Proc. ACM MobiHoc'04},
  pp.~109--120, 2004.

\bibitem{DoFrTh05}
O.~Dousse, M.~Franceschetti, and P.~Thiran, ``Information theoretic bounds on
  the throughput scaling of wireless relay networks,'' in {\em Proc. IEEE
  INFOCOM'05}, Mar. 2005.

\bibitem{DoBaTh05}
O.~Dousse, F.~Baccelli, and P.~Thiran, ``Impact of interferences on
  connectivity in ad hoc networks,'' {\em IEEE Trans. Network.}, vol.~13,
  pp.~425--436, April 2005.

\bibitem{DoFrMaMeTh06}
O.~Dousse, M.~Franceschetti, N.~Macris, R.~Meester, and P.~Thiran,
  ``Percolation in the signal to interference ratio graph,'' {\em Journal of
  Applied Probability}, vol.~43, no.~2, 2006.

\bibitem{FrDoTsTh07}
M.~Franceschetti, O.~Dousse, D.~Tse, and P.~Thiran, ``Closing the gap in the
  capacity of wireless networks via percolation theory,'' {\em IEEE Trans. on
  Information Theory}, vol.~53, no.~3, 2007.

\bibitem{KoYe07-4}
Z.~Kong and E.~M. Yeh, ``Distributed energy management algorithm for
  large-scale wireless sensor networks.'' to appear in \emph{Proc. ACM MobiHoc
  2007}, Sep. 2007.

\bibitem{KoYe08-1}
Z.~Kong and E.~M. Yeh, ``Connectivity and latency in large-scale wireless
  networks with unreliable links,'' in {\em Proc. IEEE INFOCOM'08}, Phoenix,
  AZ, April 2008.

\bibitem{KoYe08-2}
Z.~Kong and E.~M. Yeh, ``On the latency of information dissemination in mobile
  ad hoc networks,'' in {\em Proc. ACM MobiHoc'08}, Hong Kong SAR, China, May
  2008.

\bibitem{Ke87}
H.~Kesten, ``Percolation theory and first passage percolation,'' {\em Annals of
  Prob.}, vol.~15, pp.~1231--1271, 1987.

\bibitem{De03}
M.~Deijfen, ``Asymptotic shape in a continuum growth model,'' {\em Adv. in
  Applied Prob.}, vol.~35, pp.~303--318, 2003.

\bibitem{HaPeSt97}
O.~H{\"{a}}ggstr{\"{o}}m, Y.~Peres, and J.~E. Steif, ``Dynamic percolation,''
  {\em Ann. IHP Prob. et. Stat.}, vol.~33, pp.~497--528, 1997.

\bibitem{Ro95}
S.~Ross, {\em Stochastic Processes}.
\newblock New York: Wiley, second~ed., 1995.

\bibitem{Li85}
T.~Liggett, ``An improved subadditive ergodic theorem,'' {\em Annals of Prob.},
  vol.~13, pp.~1279--1285, 1985.

\bibitem{LiScSt97}
T.~M. Liggett, R.~H. Schonmann, and A.~M. Stacey, ``Domination by product
  measures,'' {\em the Ann. of Prob.}, vol.~25, no.~1.

\bibitem{Du96}
R.~Durret, {\em Probability: Theory and Examples}.
\newblock Duxbury Press, 2nd~ed., 1996.

\end{thebibliography}
\bibliographystyle{ieeetr}

\end{document}